\documentclass[twocolumn, tighten]{aastex62}
\usepackage{amsmath}
\usepackage[caption=false]{subfig}
\usepackage{xfrac}

\shorttitle{Spirals around RU Lup}
\shortauthors{Huang et al.}

\begin{document}

\title{Large-scale CO spiral arms and complex kinematics associated with the T Tauri star RU Lup}

\correspondingauthor{Jane Huang}
\email{jane.huang@cfa.harvard.edu}

\author{Jane Huang}
\affiliation{Center for Astrophysics $\mid$ Harvard \& Smithsonian, 60 Garden Street, Cambridge, MA 02138, United States of America}

\author{Sean M. Andrews}
\affiliation{Center for Astrophysics $\mid$ Harvard \& Smithsonian, 60 Garden Street, Cambridge, MA 02138, United States of America}

\author{Karin I. \"Oberg}
\affiliation{Center for Astrophysics $\mid$ Harvard \& Smithsonian, 60 Garden Street, Cambridge, MA 02138, United States of America}

\author{Megan Ansdell}
\affiliation{National Aeronautics and Space Administration Headquarters, 300 E Street SW, Washington DC 20546, United States of America}

\author{Myriam Benisty}
\affiliation{Unidad Mixta Internacional Franco-Chilena de Astronom\'{i}a, CNRS/INSU UMI 3386, Departamento de Astronom\'ia, Universidad de Chile, Camino El Observatorio 1515, Las Condes, Santiago, Chile}
\affiliation{Univ. Grenoble Alpes, CNRS, IPAG, 38000 Grenoble, France}

\author{John M. Carpenter}
\affiliation{Joint ALMA Observatory, Avenida Alonso de C\'ordova 3107, Vitacura, Santiago, Chile}

\author{Andrea Isella}
\affiliation{Department of Physics and Astronomy, Rice University, 6100 Main Street, Houston, TX 77005, United States of America}

\author{Laura M. P\'erez}
\affiliation{Departamento de Astronom\'ia, Universidad de Chile, Camino El Observatorio 1515, Las Condes, Santiago, Chile}

\author{Luca Ricci}
\affiliation{Department of Physics and Astronomy, California State University Northridge, 18111 Nordhoff Street, Northridge, CA 91130, United States of America}

\author{Jonathan P. Williams}
\affiliation{Institute for Astronomy, University of Hawaii, Honolulu, HI 96822, United States of America}

\author{David J. Wilner}
\affiliation{Center for Astrophysics $\mid$ Harvard \& Smithsonian, 60 Garden Street, Cambridge, MA 02138, United States of America}

\author{Zhaohuan Zhu}
\affiliation{Department of Physics and Astronomy, University of Nevada, Las Vegas, 4505 S. Maryland Pkwy, Las Vegas, NV, 89154, United States of America}

\begin{abstract}
While protoplanetary disks often appear to be compact and well-organized in millimeter continuum emission, CO spectral line observations are increasingly revealing complex behavior at large distances from the host star. We present deep ALMA maps of the $J=2-1$ transition of $^{12}$CO, $^{13}$CO, and C$^{18}$O, as well as the $J=3-2$ transition of DCO$^+$, toward the T Tauri star RU Lup at a resolution of $\sim0\farcs3$ ($\sim50$ au). The CO isotopologue emission traces four major components of the RU Lup system: a compact Keplerian disk with a radius of $\sim120$ au, a non-Keplerian ``envelope-like'' structure surrounding the disk and extending to $\sim260$ au from the star, at least five blueshifted spiral arms stretching up to 1000 au, and clumps outside the spiral arms located up to 1500 au in projection from RU Lup. We comment on potential explanations for RU Lup's peculiar gas morphology, including gravitational instability, accretion of material onto the disk, or perturbation by another star. RU Lup's extended non-Keplerian CO emission, elevated stellar accretion rate, and unusual photometric variability suggest that it could be a scaled-down Class II analog of the outbursting FU Ori systems. \end{abstract}


\section{Introduction} \label{sec:intro}
Mapping the distribution of dust and gas in protoplanetary disks provides insight into the conditions under which planets form. Advances in millimeter interferometry have transformed our understanding of disk structure and evolution during this past decade. High-resolution ALMA continuum observations, which trace the distribution of pebble-sized dust grains, have established a new paradigm in which young disks are often compact and organized into small-scale structures such as rings or spirals \citep[e.g.,][]{2015ApJ...808L...3A, 2018ApJ...869L..41A, 2018ApJ...869...17L, 2018ApJ...869L..43H}. 

ALMA continuum observations, though, do not provide a complete picture of the planet-forming environment. Gas constitutes the vast majority of the overall disk mass. Compared to the gas, the dust grain distribution probed by millimeter continuum emission has a smaller vertical extent due to settling and a smaller radial extent due to radial drift \citep[e.g.,][]{1977MNRAS.180...57W, 2004AA...421.1075D, 2007AA...469..213I, 2008AA...489..633P}. Hence, observations of gas tracers probe disk regions that ALMA's dust observations cannot. Disk gas is primarily composed of H$_2$, which is only directly observable in the hot inner disk atmosphere. Consequently, low-$J$ rotational transitions of CO isotopologues, which are readily mapped at millimeter wavelengths, are typically used as proxies for the gas distribution. The standard picture of gas behavior has largely been shaped by intensive CO observations of a handful of massive disks, such as those orbiting TW Hya, IM Lup, and HD 163296 \citep[e.g.,][]{2007AA...469..213I, 2012ApJ...744..162A, 2015ApJ...813...99F, 2016ApJ...832..110C, 2018ApJ...852..122H, 2018AA...609A..47P}.

\begin{deluxetable*}{cccccccc}
\tablecaption{ALMA Observing Summary \label{tab:observations}}
\tablehead{\colhead{Date} &\colhead{Configuration}&\colhead{Antennas} & \colhead{Baselines}&\colhead{Time on source}&\colhead{Amplitude}&\colhead{Bandpass}&\colhead{Phase}\\
&&&\colhead{(m)}&\colhead{(min)}}
\startdata
2018 October 2 & ACA (7-m antennas) & 11 & $9-48$ & 40 &J$1517-2422$&J$1517-2422$&J$1610-3958$ \\
2018 October 2& ACA (7-m antennas) & 12 & $9-49$ & 16 &J$1924-2914$ & J$1924-2914$ & J$1610-3958$ \\
2018 November 10& C43-5 (12-m antennas)& 45 & $15-1398$ & 33&J$1427-4206$&J$1427-4206$&J$1610-3958$\\
2018 November 13 &  C43-5 (12-m antennas)& 43 & $15-1398$&33&J$1427-4206$&J$1427-4206$&J$1610-3958$\\
2018 November 21 &  C43-5 (12-m antennas)& 43 & $15-1261$&33&J$1427-4206$&J$1427-4206$&J$1610-3958$\\
2018 December 23&  C43-2 (12-m antennas)&45&$15-500$&17&J$1517-2422$&J$1517-2422$&J$1610-3958$\\
2019 January 9&  C43-2 (12-m antennas) &44&$15-314$&16&J$1517-2422$&J$1517-2422$&J$1610-3958$\\
\enddata
\end{deluxetable*}

Increasingly, observations of other disks are demonstrating that the behavior of gas around pre-main sequence stars can differ radically from the largely axisymmetric, Keplerian disks in the most famous systems. CO maps of sources such as AB Aur, SU Aur, and DO Tau have revealed spiral arms, tails, and arcs extending well beyond the millimeter continuum emission, indicating that the planet formation environment is subject to disruptive processes that are not apparent from their compact and often symmetric dust structures \citep[e.g.,][]{2005AA...443..945P, 2012AA...547A..84T, 2019AJ....157..165A, 2020AJ....159..171F}. These discoveries motivate further investigation of the range of possible CO morphologies around young stars in order to identify the appropriate processes to include in planet formation models. 

Recent ALMA disk surveys by \citet{2018ApJ...859...21A} and \citet{2018ApJ...869L..41A} revealed large-scale, complex $^{12}$CO emission surrounding RU Lup (ICRS 15$^\text{h}$56$^\text{m}$42$^\text{s}$.311, $-37^\circ$49$'$15$\farcs$473), a $0.5\substack{+0.8\\-0.3}$ Myr old K7 star located $158.9\pm0.7$ pc away in the Lupus II cloud \citep[e.g.,][]{2017AA...600A..20A, 2018AA...616A...1G, 2018AJ....156...58B, 2018ApJ...865..157A}. One of the most extensively observed T Tauri stars, RU Lup is perhaps best known for its large and irregular spectroscopic and photometric variations from infrared to UV wavelengths \citep[e.g.,][]{1916HarCi.196....1M, 1945ApJ...102..168J, 2005AJ....129.2777H, 2013AA...560A..57G}. Because previous ALMA observations of RU Lup were not designed to recover large-scale emission, the distribution and kinematics of the gas around RU Lup were difficult to characterize with confidence. We therefore obtained new deep, multi-configuration ALMA observations of the three main CO isotopologues. The spectral setup also simultaneously targeted DCO$^+$ and SiO, which have been used in other systems as tracers of the disk midplane kinematics and of protostellar shocks, respectively \citep[e.g.,][]{1998ApJ...504L.109L,2017ApJ...843..150F}. The observational setup and data reduction are described in Section \ref{sec:observations}. Analysis of the line observations is presented in Section \ref{sec:results} and discussed in Section \ref{sec:discussion}. The conclusions are summarized in Section \ref{sec:summary}. 

\section{Observations and Data Reduction}\label{sec:observations}

\begin{deluxetable*}{ccccc}
\tablecaption{Correlator Setup \label{tab:correlatorsetup}}
\tablehead{\colhead{Target} &\colhead{Target Frequency \tablenotemark{a}}&\colhead{ACA Bandwidth}& \colhead{12-m Array Bandwidth}&\colhead{Channel Width}\\
&\colhead{(GHz)}&\colhead{(MHz)}&\colhead{(MHz)}&\colhead{(kHz)}}
\startdata
DCO$^+$ $J=3-2$ &216.1125822&62.5 &58.6&61.035 \\
SiO $J=5-4$ &217.1049800&62.5 &58.6&61.035 \\
C$^{18}$O $J=2-1$ &219.5603541&62.5 &58.6&61.035 \\
$^{13}$CO $J=2-1$ &220.3986842&62.5&58.6 &61.035 \\
$^{12}$CO $J=2-1$ &230.5380000&125.0 & 117.2&61.035 \\
Continuum window & 232.7 & 2000 &2000& 15625\\
\enddata
\tablenotetext{a}{Line rest frequencies from the Cologne Database for Molecular Spectroscopy \citep{2001AA...370L..49M, 2005JMoSt.742..215M}}
\end{deluxetable*}

ALMA Band 6 observations of RU Lup were taken using the Morita Array/Atacama Compact Array (ACA) and two configurations with the 12-meter antennas (C43-2 and C43-5). The observation dates, baseline ranges, integration times, and calibrators are given in Table \ref{tab:observations}. For each set of observations, the correlator was tuned to target the $J=2-1$ transitions of $^{12}$CO, $^{13}$CO, and C$^{18}$O, the $J=3-2$ transition of DCO$^+$, the $J=5-4$ transition of SiO, and a wide continuum window to aid in self-calibration. Details of the spectral setup are listed in Table \ref{tab:correlatorsetup}.

The raw data were first calibrated by ALMA staff with the \texttt{CASA v.5.4.0} pipeline \citep{2007ASPC..376..127M}. The same \texttt{CASA} version was used for subsequent processing and imaging following data delivery from ALMA. All execution blocks (EBs) taken in the same array configuration were self-calibrated together in the following manner: First, channels exhibiting strong line emission were flagged, and the remaining line-free channels were spectrally averaged to form a set of pseudo-continuum visibilities. The resulting  visibilities for each EB were then imaged with the \texttt{tclean} task, using the Clark CLEAN algorithm for the ACA and C43-2 data (in which the continuum emission is unresolved/marginally resolved)  and the multi-scale CLEAN algorithm for the C43-5 data (in which the continuum emission is moderately resolved). A circular mask was used for all continuum imaging. Based on 2D Gaussian fits to the continuum images, some of the EBs appeared to be offset from one another by several hundredths of an arcsecond. These small offsets were likely due to atmospheric or instrumental effects rather than proper motion, since some of the offset observations were taken only days apart, and the proper motion of RU Lup is modest ($\mu_\alpha=-11.546$, $\mu_\delta=-23.234$ mas yr$^{-1}$, \citealt{2018AA...616A...1G}).  The \texttt{fixvis} and \texttt{fixplanets} tasks were used to shift the continuum peaks to the phase center and assign a common phase  center label to all EBs, respectively. Given ALMA's systematic flux calibration uncertainty of $\sim10\%$, the visibility amplitudes from different EBs were consistent with one another at overlapping spatial frequencies. To bring the EBs into closer agreement with one another, the \texttt{gencal} task was used to rescale the C43-2 and ACA continuum visibilities to match the C43-5 continuum visibilities (the choice of flux reference is arbitrary because we do not know which observation is closest to the true flux, but our conclusions are not sensitive to flux changes on the order of a few percent). After making the small phase and flux adjustments, all of the continuum data in a given configuration were CLEANed together to yield a source model to self-calibrate the data. The ACA and C43-2 continuum observations each underwent two rounds of phase self-calibration (intervals of 60 s and 15 s for the ACA and 15 s and 10 s for C43-2) and one round of amplitude self-calibration (intervals equal to the scan length), while the C43-5 continuum underwent three rounds of phase self-calibration (intervals of 15 s, 6 s, and 6 s) and one round of amplitude self-calibration. After identifying two additional faint continuum sources in the images made separately for the C43-2 and C43-5 observations, we modified the CLEAN mask to cover these sources before combining all the configurations to produce a final continuum image. A primary beam correction was applied in \texttt{CASA} to this and all subsequent images. The nature of the additional continuum sources is discussed further in Appendix \ref{sec:contsources}. 

The self-calibration solutions derived for the spectrally-averaged continuum visibilities were then applied to the full-resolution spectral windows containing line emission. Using the \texttt{uvcontsub} task, the continuum was subtracted from the spectral line observations in the $uv$ plane. Because of the complexity of the $^{12}$CO emission, producing an image cube for analysis occurred in two parts. We produced a preliminary $^{12}$CO image with multi-scale CLEAN by manually masking the visible emission. We then searched the preliminary image for regions with significant emission. Because the noise in the image cube increases with distance from the phase center, the rms for this purpose is measured in line-free channels inside an annulus with an inner radius of $9''$ and outer radius of $10''$. We began constructing a new CLEAN mask by specifying ellipses that encompassed emission exceeding $5\times$ the rms level, with some spacing allowed between the edge of the mask and the visible emission in order to include possible lower-lying emission. When multiple $5\sigma$ regions appeared in the same channel in close proximity to one another, a single ellipse was used to enclose all of them in order to recover any faint emission that may connect them. Finally, to ensure that emission in the line wings was included, the mask was expanded to cover the regions in each channel where emission above the $5\sigma$ level appears in an adjacent channel. The resulting CLEAN mask, which was used for a new round of imaging (beginning again from a dirty image and not from the preliminary image), is shown in Appendix \ref{sec:maskdescription}. After CLEANing down to a threshold of 15 mJy beam$^{-1}$, the mask in the channels from 3.5 to 6.5 km s$^{-1}$ was expanded to the width of the primary beam in order to encompass the cloud emission. Deconvolution then proceeded until a threshold of 5 mJy beam$^{-1}$ was reached. For the more compact $^{13}$CO and C$^{18}$O emission, CLEAN masks covering the area of the primary beam were used for the cloud-contaminated central channels, and circular masks with a radius of $2''$ were used for the other channels with disk emission. The weaker DCO$^+$ emission was imaged with a Gaussian taper to improve sensitivity. Given the larger synthesized beam, its CLEAN mask was set to a radius of $3''$. Because we are only concerned with emission near the phase center for these three lines, the image cube rms is measured inside an $8''\times8''$ box at the phase center. The image properties are listed in Table \ref{tab:imageproperties}. 

\begin{deluxetable*}{ccccccc}
\tablecaption{Imaging Summary\label{tab:imageproperties}}
\tablehead{
&\colhead{Briggs parameter}&\colhead{Synthesized beam}&\colhead{Peak $I_\nu$}&\colhead{RMS noise\tablenotemark{c}}&\colhead{Integrated Flux\tablenotemark{d}}\\
&&(arcsec $\times$ arcsec ($^\circ$))&(mJy beam$^{-1}$)&(mJy beam$^{-1}$)&\colhead{(Jy km s$^{-1}$)}}
\startdata
$^{12}$CO $J=2-1$ & 0.5 &$0.32\times0.31$ $(-57.5)$& 307&2.5\tablenotemark{a} (1.7\tablenotemark{b})& $\sim20$ \\
$^{13}$CO $J=2-1$ & 0.5 & $0.32\times0.32$ $(-56.6)$ &93 &2.0\tablenotemark{b} &$1.2\pm0.2$\\
C$^{18}$O $J=2-1$ & 0.5&$0.33\times0.32$ $(-47.2)$& 47& 1.6\tablenotemark{b} &$0.34\pm0.03$ \\
DCO$^+$ $J=3-2$ & 2.0 & $0.50\times0.48$ $(69.1)$& 12 & 1.3\tablenotemark{b}&$0.054\pm 0.016$ \\
\enddata
\tablenotetext{a}{Measured in an annulus centered on the disk, with an inner radius of $9''$ and outer radius of $10''$. }
\tablenotetext{b}{Measured in an $8''\times8''$ box with the same center as the disk.}
\tablenotetext{c}{Measured in channels with $dv=0.25$ km s$^{-1}$.}
\tablenotetext{d}{Only statistical uncertainties are quoted in the table. The systematic flux calibration uncertainty contributes another $10\%$.}
\end{deluxetable*}

No SiO emission is detected down to an rms level of 1.6 mJy beam$^{-1}$ and a tapered synthesized beam of $0\farcs79\times0\farcs74$ $(77\fdg0)$. Given that SiO has generally been observed in protostellar outflows rather than disks \citep[e.g.,][]{1998ApJ...504L.109L, 2019AA...632A.101T}, we do not have suitable constraints on the likely emitting area of SiO. Therefore, we do not seek to estimate a flux upper limit for SiO or apply weak-line detection techniques that take advantage of the known Keplerian velocity field of the disk \citep[e.g.,][]{2016ApJ...832..204Y, 2018AJ....155..182L}.

\section{Results \label{sec:results}}
\begin{figure*}[h]
\begin{center}
\includegraphics{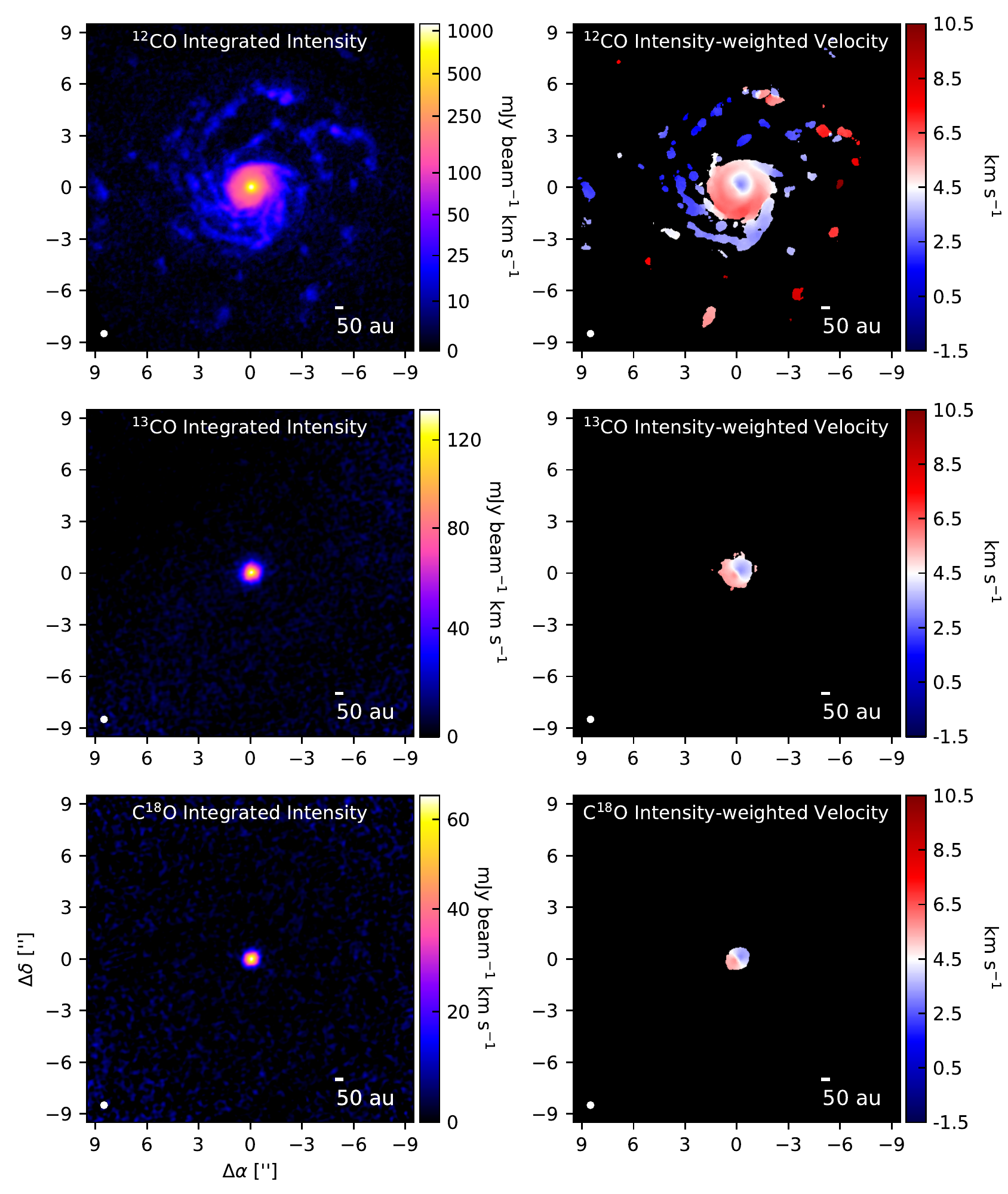}
\end{center}
\caption{Integrated intensity and intensity-weighted velocity maps of $^{12}$CO, $^{13}$CO, and C$^{18}$O $J=2-1$ toward RU Lup. An asinh stretch is used on the $^{12}$CO integrated intensity color scale in order to show the weak emission at large distances from the phase center, while $^{13}$CO and C$^{18}$O are shown on linear scales. The synthesized beam is plotted in the lower left corner of each panel. Offsets from the phase center are marked on the axes. \label{fig:COmomentmaps}}
\end{figure*}

\subsection{Overview of CO isotopologue emission morphology}
RU Lup's CO emission exhibits a great degree of morphological and kinematic complexity. While Appendix \ref{sec:maskdescription} shows that $^{12}$CO is detected at the $3\sigma$ level over a wide velocity range ($-5.5$ to $10.75$ km s$^{-1}$), the emission at any given spatial location is detected over a much narrower range. To produce an integrated intensity map of a disk without degrading the signal-to-noise ratio at a given spatial location by including channels at velocities where emission is known to be absent, it has become increasingly common to apply a ``Keplerian mask'' to the image cube before performing the integration \citep[e.g.,][]{2017AA...606A.125S, 2018ApJ...859...21A}. This strategy is not suitable for RU Lup, given the highly non-Keplerian emission pattern of $^{12}$CO. Instead, we use the CLEAN mask shown in Appendix \ref{sec:maskdescription}  to select pixels for inclusion. Because cloud contamination is visible between 3.75 and 6.5 km s$^{-1}$, pixels in the image cube that fall below the 1$\sigma$ level are also excluded when calculating the integrated intensity map in order to minimize distortion of RU Lup's CO structures. Intensity-weighted velocity maps are computed in a similar manner, except the threshold for including pixels in the calculation is set to $5\sigma$ because higher moment maps are more readily distorted by noise outliers. 

The $^{13}$CO and C$^{18}$O lines also exhibit cloud contamination, although to a lesser extent than $^{12}$CO. Integrated intensity maps are produced by collapsing the cube between 1.75 and 7.25 km s$^{-1}$, the range of channels where emission above the 3$\sigma$ level is detected. After ascertaining that no extended emission structures are detected at the $3\sigma$ level, intensity-weighted velocity maps are computed by applying a circular mask with a radius of $2''$ and excluding pixels below the 5$\sigma$ level. The maps for the CO isotopologues are shown in Figure \ref{fig:COmomentmaps}.

The integrated intensity maps reveal four major gas components around RU Lup:
\begin{enumerate}
\item A Keplerian disk extending to a radius of $\sim120$ au. The Keplerian rotation is most clearly traced by the intensity-weighted velocity map of C$^{18}$O. The systemic velocity appears to be $\sim4.5$ km s$^{-1}$, with emission northwest of the star being blueshifted relative to the systemic velocity and emission to the southeast being redshifted relatively. $^{13}$CO follows a similar rotation pattern, but with some mild asymmetries at its edge. The rotation pattern traced by C$^{18}$O can also be seen in the inner regions of the $^{12}$CO velocity field, but $^{12}$CO is dominated by non-Keplerian emission. 
\item An envelope-like structure extending from the Keplerian disk to a radius of $\sim260$ au. Here, we use ``envelope-like" in the sense that it is kinematically distinct from but wholly encircles the Keplerian disk. This terminology is chosen in order to refer to this structure concisely in the remainder of this paper, but does not necessarily mean that this structure is an envelope as traditionally understood in star formation (i.e., material infalling onto a protostar following gravitational collapse of the parent cloud). The envelope-like structure is most clearly visible in $^{12}$CO. The mild outer asymmetries in the $^{13}$CO velocity map trace parts of the envelope-like structure. 
\item At least five clumpy spiral arms extending from the edge of the envelope-like structure, with some stretching up to a projected distance of $\sim1000$ au from RU Lup. The spiral arms are largely blueshifted relative to the systematic velocity and are only detected in $^{12}$CO.
\item Multiple localized $^{12}$CO intensity enhancements outside the spiral arms. At the sensitivity of our observations, they do not appear to be connected to the main disk structure or to the spiral arms. For brevity, we refer to these localized intensity enhancements as ``clumps.'' The clumps are largely redshifted relative to the systemic velocity. They are not detected in $^{13}$CO or C$^{18}$O. 
\end{enumerate}  

The integrated fluxes of the CO isotopologues are reported in Table \ref{tab:imageproperties}. The $^{12}$CO flux, which is estimated simply by summing over the clipped, masked integrated intensity map,  is reported for completeness but should be treated with caution. The prominent cloud contamination near the systemic velocity and the large extent and irregularity of the $^{12}$CO emission make it difficult to estimate a flux uncertainty. In addition, the $^{12}$CO emission may still suffer from some spatial filtering even outside the cloud-contaminated channels. Using the ALMA Technical Handbook's definitions\footnote{\url{https://almascience.nrao.edu/documents-and-tools/cycle7/alma-technical-handbook/view}} of maximum recoverable scale (MRS), the MRS defined by the shortest baseline $L_\text{min}$ ($\theta_\text{MRS}\approx \sfrac{0.6\lambda_\text{obs}}{L_\text{min}}$) is $20''$ and the more conservative MRS defined by the 5th percentile of the baselines ($\theta_\text{MRS}\approx \sfrac{0.983\lambda_\text{obs}}{L_\text{5}}$) is $7''$. Within individual channels, the spiral arms have a spread of $\sim11''$. 

The $^{13}$CO and C$^{18}$O integrated fluxes are measured from the integrated intensity maps within $3''$ diameter apertures. The aperture size is chosen based on examination of the radial extent of $^{13}$CO emission above the $3\sigma$ level in channels that are not cloud-contaminated. The flux uncertainties are estimated by measuring fluxes within a $3''$ diameter aperture at 100 random off-source positions in the integrated intensity maps and taking the standard deviation. We note that even though the S/N of $^{13}$CO is higher than that of C$^{18}$O in individual channels, the flux uncertainty for $^{13}$CO is much higher because of larger contributions from cloud contamination to off-source positions in the integrated intensity map. The $^{13}$CO and C$^{18}$O fluxes are consistent with previous RU Lup observations reported in \citet{2018ApJ...859...21A}.

\begin{figure*}
\begin{center}
\includegraphics{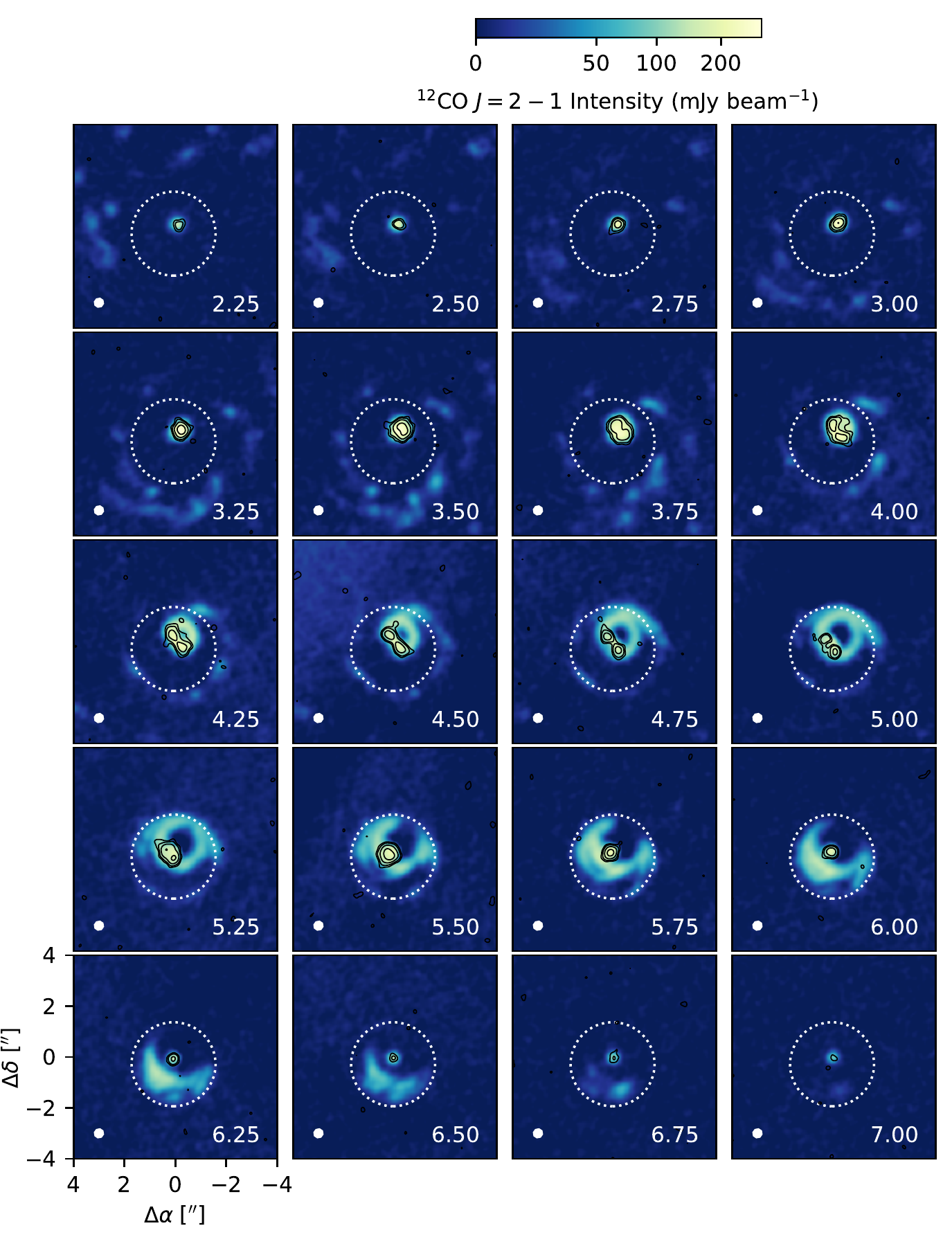}
\end{center}
\caption{Insets of $^{12}$CO channel maps near the systemic velocity, highlighting the inner regions of the RU Lup system. C$^{18}$O contours are drawn at [3, 5, 10, 20]$\sigma$ to show the rotation pattern of the Keplerian disk. The white dotted circle marks the outer boundary of the ``envelope-like'' structure at a radius of $\sim260$ au. The synthesized beam is plotted in the lower left corner of each panel. Offsets from the phase center are marked on the axes. \label{fig:inner12CO}}
\end{figure*}

\begin{figure*}
\begin{center}
\includegraphics{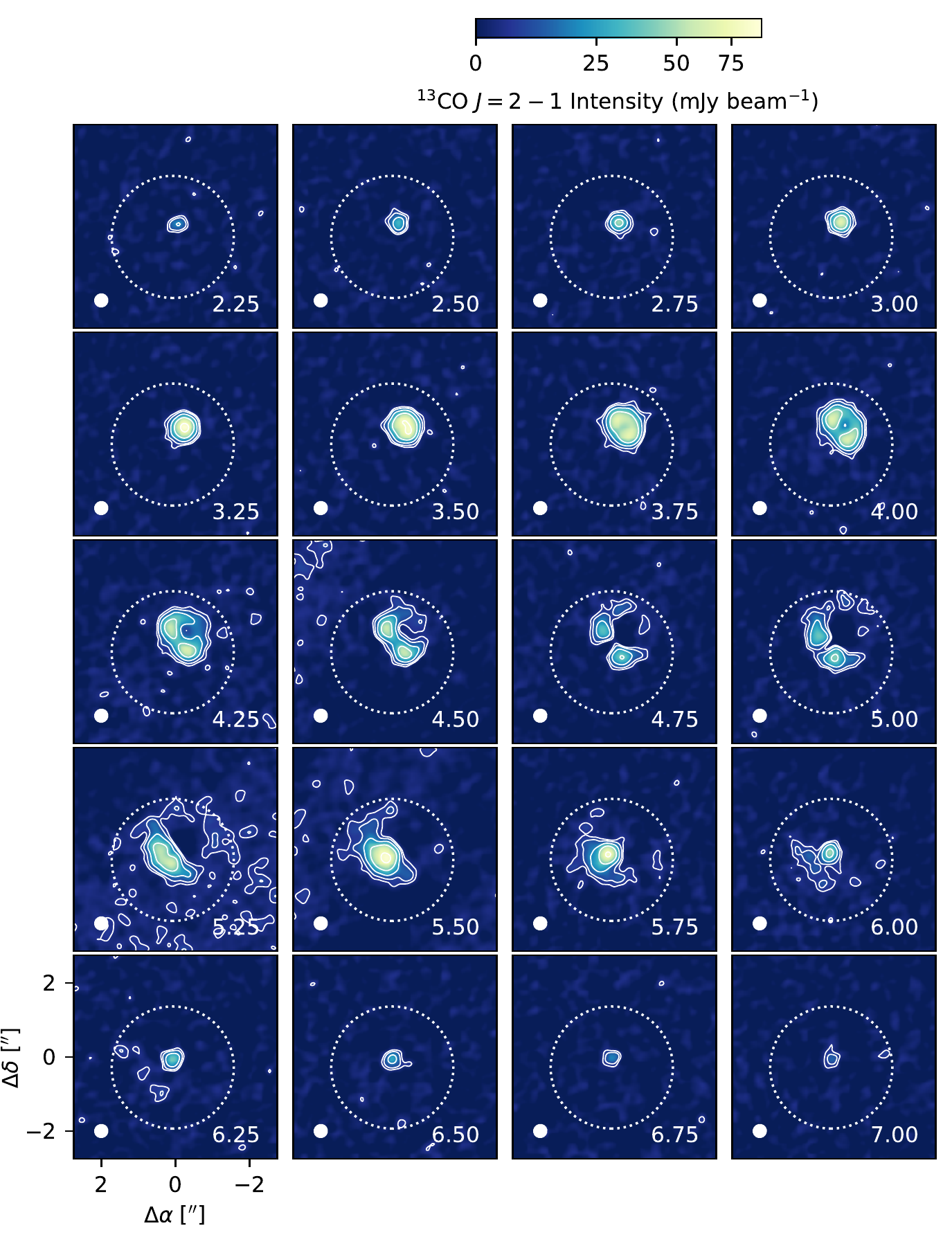}
\end{center}
\caption{$^{13}$CO channels near the systemic velocity. The $^{13}$CO contours are drawn at [3, 5, 10, 20, 40]$\sigma$. The white dotted circle marks the outer boundary of the ``envelope-like'' structure in $^{12}$CO emission (see Figure \ref{fig:inner12CO}). The large-scale diffuse emission visible in some channels is due to cloud contamination. The synthesized beam is plotted in the lower left corner of each panel. Offsets from the phase center are marked on the axes. \label{fig:inner13CO}}
\end{figure*}

\subsubsection{The inner regions of the RU Lup system\label{sec:innerregions}}

To highlight the structure of the inner regions of the RU Lup system, Figure \ref{fig:inner12CO} presents $8''\times8''$ maps of channels near the systemic velocity for $^{12}$CO, with C$^{18}$O contours overlaid to show the rotation of the Keplerian disk. The ``envelope-like'' structure is visible from $\sim4.25$ to 7.00 km s$^{-1}$. From 4.25 to 5.50 km s$^{-1}$, the structure manifests as a loop of emission emerging (in projection) from the northwest side of the Keplerian disk and growing in size with increasing redshift. The disk and loop are surrounded by a clumpy emission ring with a radius of $\sim$260 au (marked with a dotted white circle). From 5.50 km s$^{-1}$ to 7 km s$^{-1}$, the clumpy emission ring is no longer visible, but the loop northwest of the disk transitions to crescent-like emission southeast of the Keplerian disk. This crescent-like emission lies entirely within the region previously traced by the clumpy emission ring. 

The movement of emission from northwest to southeast of RU Lup is in the same direction as the Keplerian disk, suggesting that the ``envelope-like'' structure is rotating. However, the bulk of the emission is redshifted relative to the systemic velocity (identified as $\sim4.5$ km s$^{-1}$ based on the characteristic Keplerian ``hourglass'' emission pattern of C$^{18}$O at this velocity). This implies that the structure's bulk motion along the line of sight is different from that of the RU Lup disk. If the structure is positioned between the observer and RU Lup, this would suggest that it is infalling. If the structure lies behind RU Lup relative to the observer, then the kinematics would indicate that the structure is moving away from RU Lup, perhaps as part of an outflow or wind. 

The envelope-like structure is also marginally visible in the $^{13}$CO channel maps, as shown in Figure \ref{fig:inner13CO}. Portions of the loop northwest of the disk are visible from 4.25 km s$^{-1}$ to 5.25 km s$^{-1}$ and the crescent southwest of the disk is visible from 5.50 km s$^{-1}$ to 6.25 km s$^{-1}$.

Given that the envelope-like structure is most prominent in the cloud-contaminated channels, one may question whether it is part of the RU Lup system or simply cloud emission that coincides with RU Lup in projection. We consider the latter scenario unlikely for a couple reasons. First, the ``emission loop'' observed between 4.25 and 5.50 km s$^{-1}$ connects smoothly with the outer edge of the disk emission traced by C$^{18}$O (Figure \ref{fig:inner12CO}), suggestive of a physical relationship between the structures. Second, while \citet{2018ApJ...859...21A} find that a number of disks in Lupus exhibit cloud contamination in $^{12}$CO, they note that the presence of the envelope-like structure (which they refer to as a possible outflow) is distinct to RU Lup. This suggests at least that cloud contamination does not commonly produce features resembling RU Lup's envelope-like structure.

\begin{figure*}
\begin{center}
\includegraphics{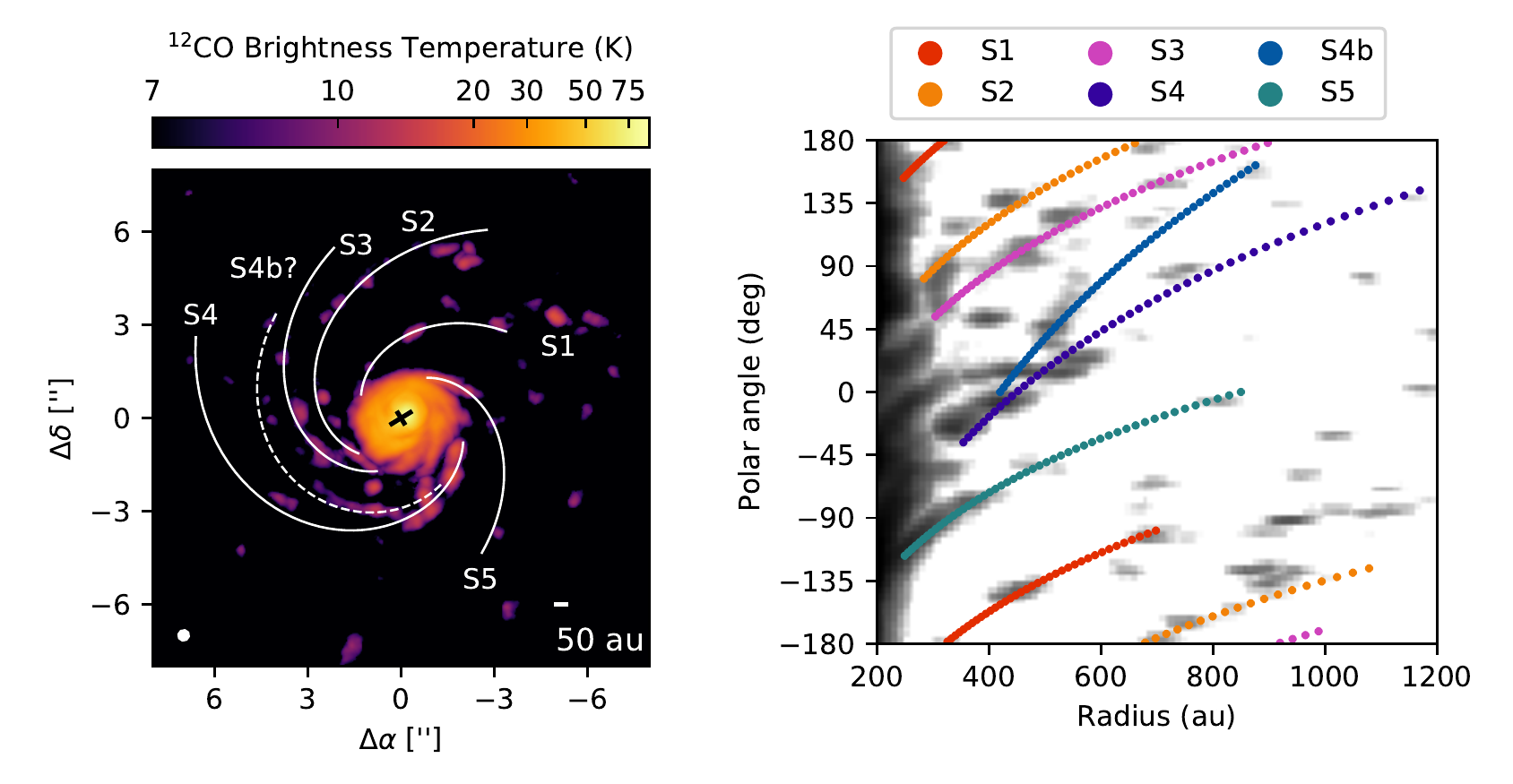}
\end{center}
\caption{\textit{Left}: Peak brightness temperature map of $^{12}$CO toward RU Lup with the spiral arms labeled. The black cross marks the position of RU Lup's continuum peak. The longer line marks RU Lup's projected major axis. The synthesized beam is plotted in the lower left corner of the panel. Offsets from the phase center are marked on the axes. \textit{Right}: The $^{12}$CO peak brightness temperature map deprojected and replotted as a function of disk radius and polar angle. \label{fig:spiralsannotated}}
\end{figure*}

\subsubsection{Properties of RU Lup's CO spiral arms\label{sec:spiralarms}}
The clumpiness and close proximity of RU Lup's CO spiral arms make it challenging to identify them individually in the integrated intensity map. To isolate the brightest parts of the spiral arms, we make a peak brightness temperature map. Using non-continuum subtracted $^{12}$CO measurement sets (so as not to underestimate temperatures at the disk center), we produce three sets of image cubes with channelizations shifted 0.1 km s$^{-1}$ apart from one another. Peak intensity maps are generated from each cube, median-stacked, and then converted to brightness temperatures using the full Planck equation. The peak intensities are measured only within the CLEAN mask to exclude emission from cloud contamination, and the map is clipped at the $5\sigma$ level to minimize distortion from noise outliers. The median-stacking procedure is necessary to mitigate the artifacts that arise from channelization \citep[e.g.,][]{2014ApJ...785L..12C}. Using the inclination ($18\fdg8$) and position angle (121$^\circ$) measured for RU Lup in \citet{2018ApJ...869L..43H}, the peak brightness temperature map is then deprojected and replotted as a function of polar angle and disk radius. 

Figure \ref{fig:spiralsannotated} shows the $^{12}$CO brightness temperature map and polar plot, with logarithmic spiral curves ($R(\theta) = R_0 e^{b\theta}$) overlaid for visual guidance. Due to the irregularity of the emission, the parameters for the proposed spirals were adjusted manually in order to identify a reasonable approximation to the spiral morphology. The parameter values and the associated pitch angles ($\arctan \left |b\right|$) are listed in Table \ref{tab:spirals}. 

The logarithmic spiral form is chosen as a simple way of describing the large-scale behavior of the spiral arms, but it does not fully capture the nuances of their geometry. It is sometimes ambiguous which emission features should be assigned to the spirals, and other reasonable spirals may also be drawn. S4 and S5 are drawn such that they terminate at emission clumps that are well-separated from the rest of the spiral emission, but appear to fall on the same arc and share similar kinematics (see Figure \ref{fig:COmomentmaps}). Meanwhile, the outer tips of S1 and S2 are contiguous with emission features that are redshifted by several km s$^{-1}$ compared to the rest of the spiral arms, so it is unclear whether the redshifted emission features are part of the arms. S1 appears to fork at $R\sim700$ au, and S4 appears to fork at $\sim400$ au, leading to an emission arc we label S4b. S4/S4b are particularly challenging to describe because there also appears to be a CO filament located at a polar angle of $0^\circ$ that emerges from the envelope-like structure and then merges into S4/S4b. Furthermore, the spiral geometries are computed under the assumption that they lie in the same plane as RU Lup's disk. However, given that their kinematics are quite different from the disk, it is plausible that they are not coplanar.

\begin{deluxetable}{cccc}
\scriptsize
\tablecaption{Proposed Spirals\label{tab:spirals}}
\tablehead{
\colhead{ID}&\colhead{$R_0$} &\colhead{ b} & \colhead{Pitch angle} \\
&\colhead{(au)} & &\colhead{(Degrees)}}
\startdata
S1&  57 & 0.55 & 29 \\
S2 & 140 & 0.5 & 27 \\
S3 & 190 & 0.5 & 27 \\
S4 & 450 & 0.38 & 21 \\
S4b & 420 & 0.26 & 15 \\
S5 & 850 & 0.6 & 31 \\
\enddata
\end{deluxetable}

\begin{figure*}[!p]
\begin{center}
\includegraphics{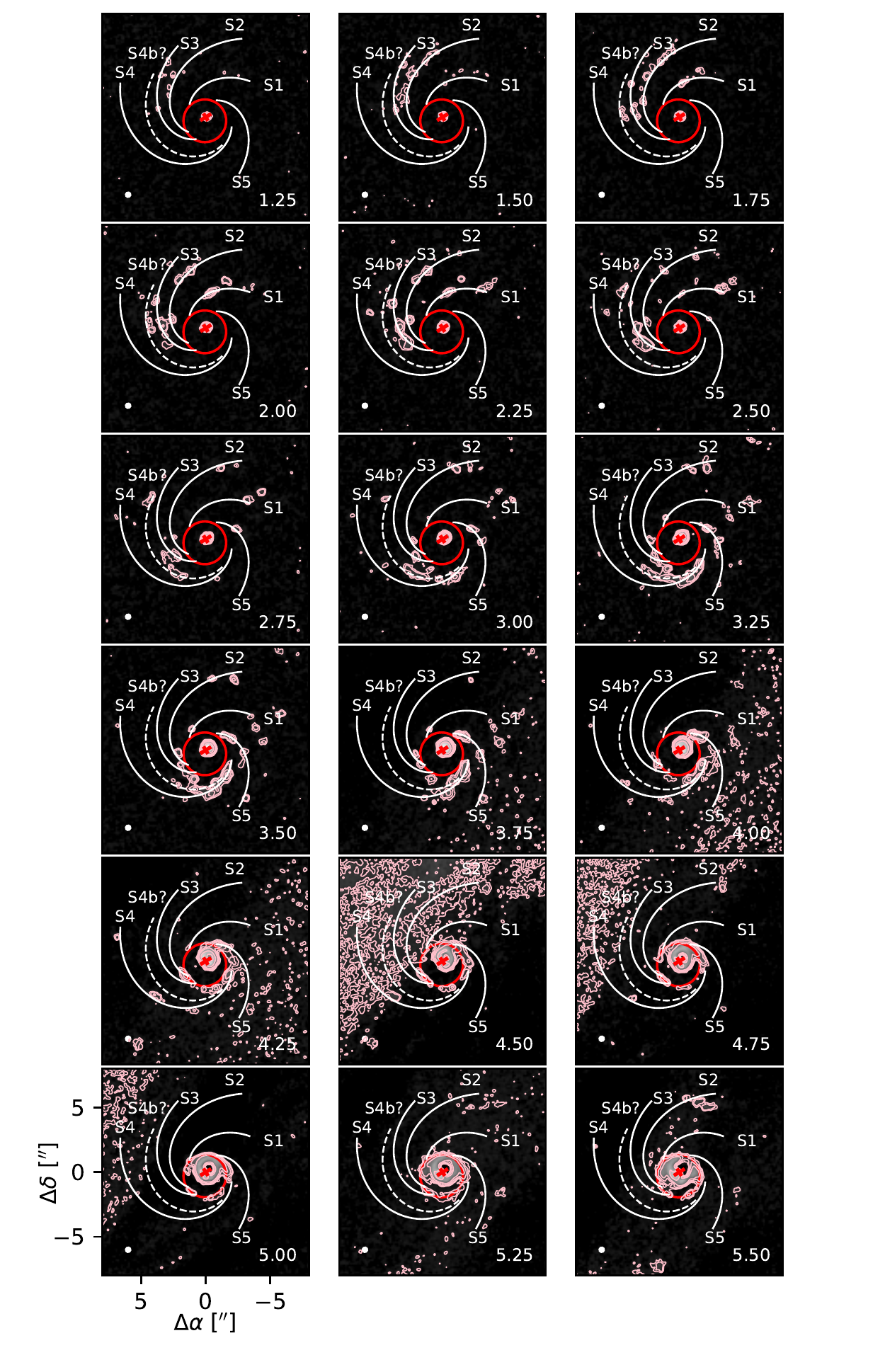}
\end{center}
\caption{$^{12}$CO channel maps with the spiral locations labeled. The contours are drawn at the [3, 5, 10, 20, 50]$\sigma$ level. The red circle marks the location of the clumpy emission ring described in Section \ref{sec:innerregions}. The red cross marks the position of the RU Lup disk's continuum peak, with the longer line marking RU Lup's projected major axis. The synthesized beam is plotted in the lower left corner of each panel. Offsets from the phase center are marked on the axes. Note that cloud contamination is present in the panels ranging from 3.75 to 5.50 km s$^{-1}$ (as well as at additional velocities not shown). \label{fig:spiralchanmaps}}
\end{figure*}

To examine the kinematics of the spiral arms in more detail, we overlay the model spiral curves on the $^{12}$CO channel maps in Figure \ref{fig:spiralchanmaps}. The spirals are detected from $\sim1.25-4.75$ km s$^{-1}$, i.e., they are largely blueshifted from the systemic velocity of $\sim4.5$ km s$^{-1}$. The inner tips of the spiral arms terminate at the clumpy emission ring that marks the outer boundary of the ``envelope-like'' structure described in Section \ref{sec:spiralarms}. This spatial relationship again suggests that the envelope-like structure is not simply foreground cloud emission. In all channels, the spiral arm emission is clumpy. The spiral arms are detected at $\sim10\sigma$ in individual channels, but we observe voids along the spiral arms where no emission is detected above the $3\sigma$ level. This implies that intensity varies locally by at least a factor of 3 along the arms. With only a single transition, it is not clear whether these intensity variations are due to rapidly changing column densities or temperatures. Cloud contamination does not account for the clumpiness, given that the clumpiness also appears at velocities offset from where cloud emission is observed ($\sim3.75-6.5$ km s$^{-1}$). Spatial filtering may exaggerate the clumpiness of the emission but is unlikely to be solely responsible. The non-cloud-contaminated channels do not exhibit the imaging artifacts commonly associated with spatial filtering, such as striping or negative CLEAN bowls around the emission (as expected, these artifacts are clearly visible in some of the cloud-contaminated channels). We also imaged the observations from the ACA and combined 12-m configurations separately, and found that the CO fluxes were comparable (within $\sim10\%$) in the non-cloud-contaminated channels. If the C43-2 configuration had substantially resolved out RU Lup's large-scale emission, we would expect the CO fluxes measured with the ACA to be much higher.

The channel maps and intensity-weighted velocity map show that the spiral emission tends to be more blueshifted at larger projected distances from RU Lup, indicating that the gas in the outer regions of the spirals is moving faster relative to RU Lup along the line of sight compared to the inner regions.  Gas velocities increasing with distance from the star is qualitatively consistent with expectations for outflows from young stars \citep[e.g.,][]{1991ApJ...370L..31S} but opposite from predictions of standard infall models \citep[e.g.,][]{1976ApJ...210..377U}. That said, a major difficulty in interpreting the kinematics is that the three-dimensional distribution of gas is unclear, so the relative contributions of radial and rotational motion to the observed line-of-sight velocities are ambiguous. Given the positions of the spiral arms, one might alternatively describe the spiral kinematics in terms of a velocity gradient with increasing blueshift from southwest to northeast of RU Lup. The direction of this apparent gradient is perpendicular to that of both the Keplerian disk and the ``envelope-like'' structure.

\begin{figure*}
\begin{center}
\includegraphics{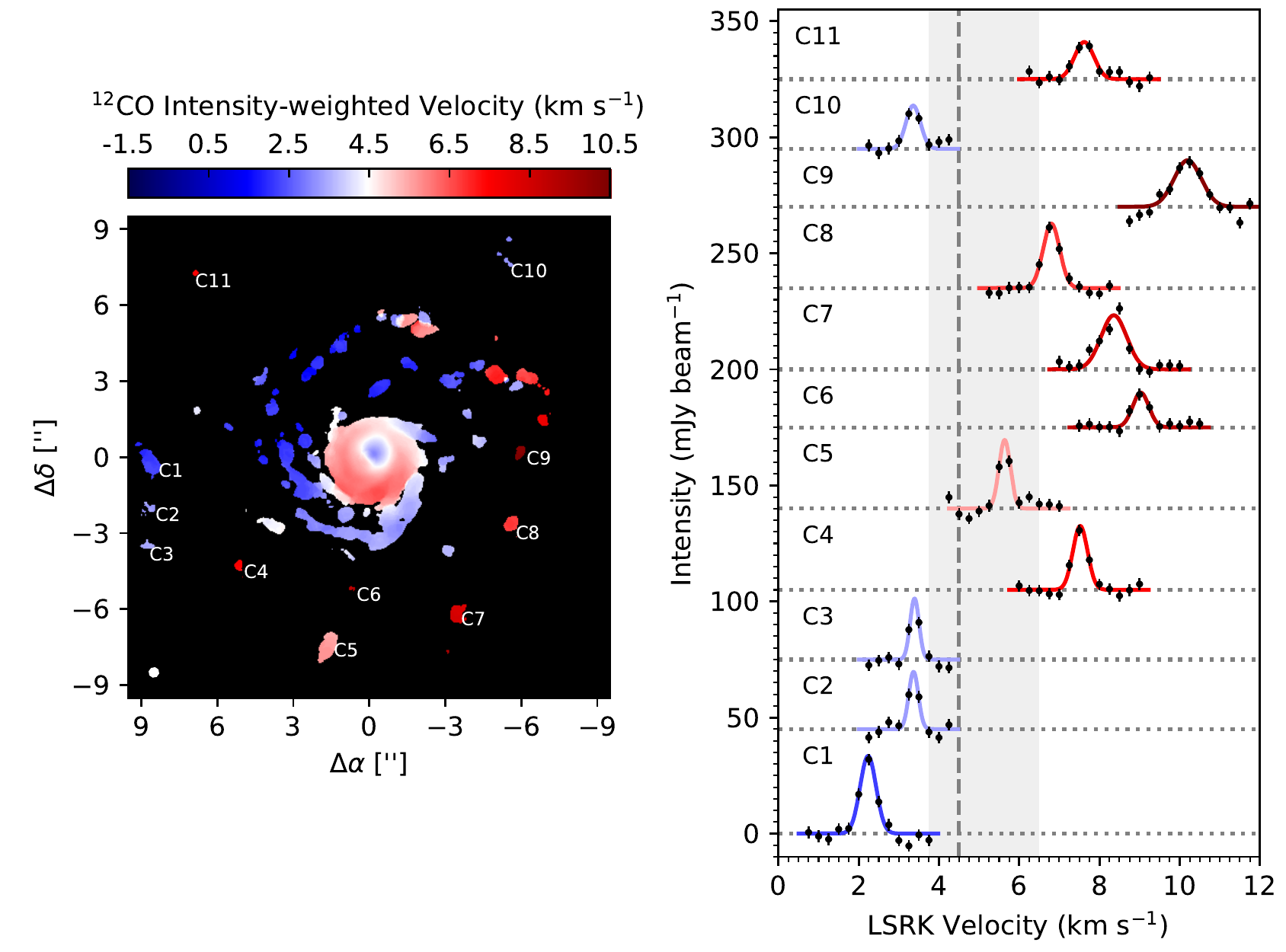}
\end{center}
\caption{\textit{Left}: The $^{12}$CO intensity-weighted velocity map with the clumps outside the spiral arms labeled. \textit{Right}: Spectra extracted from the peaks of the clumps (black points) and best-fit model spectra (color-coded by the central velocity). Dotted horizontal lines show the zero-intensity level for each clump. Error bars show the $1\sigma$ uncertainty in the intensity in each channel (excluding the 10\% flux calibration uncertainty). The dashed vertical line marks RU Lup's systemic velocity ($\sim4.5$ km s$^{-1}$). The velocity range over which cloud contamination is visible is shaded in gray. \label{fig:clumpspectra}}
\end{figure*}

\subsubsection{Properties of CO emission clumps outside the spiral arms}
A feature in $^{12}$CO emission is identified as a ``clump'' outside the spiral arms if it meets the following criteria:
\begin{itemize}
\item Emission exceeds the $5\sigma$ level (where $\sigma$ is the RMS noise level reported in Table \ref{tab:imageproperties}) in at least one channel, without detection of any emission at the 3$\sigma$ level connected to a larger structure, such as the spiral arms, the Keplerian disk, or ambient cloud emission. 
\item Emission exceeding the $3\sigma$ level can be found in at least one adjacent channel at the same spatial location, to confirm that the emission is not merely a noise spike or imaging artifact in a single channel. 
\item If the emission appears at the same velocity as the spiral arms, it does not fall along an arc traced by the spiral arms. 
\end{itemize}
The criteria are chosen to be conservative, thus disfavoring the identification of clumps that have line widths narrower than the channel width of $dv=0.25$ km s$^{-1}$, occur primarily at velocities affected by cloud contamination, or appear in close proximity to the spiral arms. Thus, these clump identifications are not exhaustive. 

The clumps are sufficiently numerous such that it seems unlikely that the emission all originates from foreground or background objects that are unrelated to RU Lup. The clumps do not coincide with any sources in the \textit{Gaia} database \citep{2018AA...616A...1G}. The background star detected in VLT-SPHERE $H$-band images of RU Lup by \citet{2018ApJ...863...44A} does not coincide with any of the clumps either. 

The clumps are numbered starting east of RU Lup and going counterclockwise in the intensity-weighted velocity map (Figure \ref{fig:clumpspectra}). The location of each clump is defined to be at the pixel corresponding to the maximum value in the peak brightness temperature map. The projected offset and distance are computed with respect to the continuum peak of the RU Lup disk, which is presumed to coincide with the position of the star since the dust emission appears to be symmetric even at very high resolution \citep[e.g.,][]{2018ApJ...869L..42H}. The uncertainty in the position is taken to be the standard deviation of the Gaussian synthesized beam. These quantities are reported in Table \ref{tab:clumpproperties}.

\begin{deluxetable*}{cccccccc}
\scriptsize
\tablecaption{Clump properties\label{tab:clumpproperties}}
\colnumbers
\tablehead{
\colhead{ID}&\colhead{Position \tablenotemark}&\colhead{Offset}&\colhead{T$_B$}&\colhead{Velocity}&\colhead{Line width}&\colhead{Size}&\colhead{Flux}\\
&\colhead{(arcsec, arcsec)}&\colhead{(au)}&\colhead{(K)}&\colhead{(km s$^{-1}$)}&\colhead{(km s$^{-1}$)}& (au $\times$ au) &\colhead{(mJy km s$^{-1}$)}}
\startdata
C1 & ($8.56\pm0.14$, $-0.28\pm0.14$) & $1360\pm30$ &$12.7\substack{+0.8\\-0.7}$ &$2.23\pm0.02$ &$0.47\substack{+0.05\\-0.04}$ & $(180\pm20)\times(75\pm11)$& $108\pm13$\\
C2\tablenotemark{a} & ($8.68\pm0.14$, $-2.04\pm0.14$) & $1420\pm30$ &$10.4\substack{+3\\-1.6}$& $3.37\pm0.02$ &$0.28\pm0.1$&$(150\pm30)\times(120\pm30)$&$50\pm11$\\
C3\tablenotemark{a} & ($8.92\pm0.14$, $-3.60\pm0.14$) & $1530\pm30$ &$10.8\substack{+3\\-1.7}$&$3.39\substack{+0.03\\-0.02}$ & $0.26\pm0.1$&$(130\pm30)\times(90\pm30)$&$42\pm11$\\
C4 & ($5.20\pm0.14$, $-4.28\pm0.14$) & $1070\pm30$ &$11.2\substack{+0.8\\-0.7}$& $7.52\pm0.02$ &$0.43\substack{+0.06\\-0.05}$& $(150\pm30)\times(54\pm13)$ & $51\pm9$ \\
C5\tablenotemark{a}  &  ($1.64\pm0.14$, $-7.40\pm0.14$) &$1210\pm30$ &$11.7\substack{+3\\-1.3}$& $5.63\pm0.02$&$0.33\substack{+0.09\\-0.11}$&$(200\pm20)\pm(109\pm12)$&$102\pm11$\\
C6 & ($0.76\pm0.14$, $-5.20\pm0.14$) & $840\pm30$ &$7.8\pm0.8$&$9.03\pm0.05$&$0.47\substack{+0.11\\-0.09}$&$(80\pm20)\times(60\pm20)$&$25\pm6$ \\  
C7 & ($-3.36\pm0.14$, $-6.16\pm0.14$) & $1120\pm30$ &$10.0\pm0.6$& $8.36\substack{+0.03\\-0.04}$ &$0.75\pm0.09$ &$(141\pm16)\times(109\pm12)$&$116\pm12$\\
C8\tablenotemark{a} & ($-5.52\pm0.14$, $-2.76\pm0.14$) & $980\pm30$ &$11.2\substack{+0.08\\-0.07}$&$6.80\pm0.02$ &$0.49\pm0.06$&$(119\pm17)\times(104\pm15)$&$57\pm8$\\
C9 & ($-5.96\pm0.14$, $0.20\pm0.14$)& $950\pm30$ &$9.2\pm0.5$& $10.22\pm0.04$ & $0.79\pm0.09$&$(85\pm13)\times (40\pm10)$&$42\pm5$ \\
C10\tablenotemark{a} & ($-5.56\pm0.14$, $7.60\pm0.14$) & $1500\pm30$ &$8.8\substack{+1.4\\-1.0}$&$3.36\pm0.03$ & $0.43\substack{+0.13\\-0.12}$&$(260\pm60)\times(110\pm30)$ &$73\pm16$\\
C11 & ($6.88\pm0.14, 7.20\pm0.14$)& $1580\pm30$ &$8.1\pm0.8$ & $7.62\pm0.04$ & $0.56\substack{+0.15\\-0.11}$ &$(130\pm30)\times(50\pm20)$&$38\pm9$ 
\enddata
\tablecomments{(1) Identifier for clump. (2) Position relative to RU Lup. The coordinate corresponds to the offset east and north of RU Lup, respectively. (3) Projected offset from RU Lup. (4) Peak brightness temperature of Gaussian fit to clump spectrum. The uncertainties quoted are statistical; systematic flux calibration errors contribute another $\sim10\%$ to the uncertainties. (5) LSRK velocity corresponding to the peak of the Gaussian fit to the clump spectrum. (6) FWHM of Gaussian fit to clump spectrum. (7) FWHM of two-dimensional Gaussian fit to the clump in its integrated intensity image.  (8) Integrated flux measured from two-dimensional Gaussian fit to the clump in its integrated intensity image. Systematic flux calibration errors contribute another $\sim10\%$ to the quoted errors.}
\tablenotetext{a}{The measured properties of the clump are affected by cloud contamination}
\end{deluxetable*}

At each of these clump peak pixel locations, we extract a spectrum from the $^{12}$CO image cube. To estimate the peak intensity $I_\text{peak}$, central velocity $v_0$, and line standard deviation $\sigma_v$, we assume that the underlying spectrum is described by a Gaussian of the form 
\begin{equation}
I(v) = I_\text{peak}\exp\left(-\frac{(v-v_0)^2}{2\sigma_v^2} \right),
\end{equation}
 where $v$ is the LSRK velocity. To account for instrumental broadening, we convolve the Gaussian with a tophat filter with the same width as the image cube channels. The spectrum is then sampled at the same velocities as the image cube, generally over a range of $\sim3$ km s$^{-1}$ around the clump peak. This range is sometimes truncated when the line wings are affected by cloud contamination. The likelihood function is taken to be Gaussian, and the noise in each channel is assumed to be independent. These approximations are reasonable in the regime where the channel width of the line cube is at least a few times larger than the native resolution of the data following Hanning smoothing ($\sim0.09$ km s$^{-1}$ for the $^{12}$CO line).\footnote{\url{https://safe.nrao.edu/wiki/pub/Main/ALMAWindowFunctions/Note_on_Spectral_Response_V2.pdf}} For contexts in which Hanning smoothing is expected to have a significant impact on the inferred line widths and high precision measurements are desired, \citet{2016ApJ...831...16L} describe a more detailed method to account for ALMA's spectral response. Broad uniform priors are specified: $0<I_\text{peak}<1$ Jy beam$^{-1}$, $0.05<\sigma_v<3$ km s$^{-1}$, and $v_0$ within the velocity range of the spectrum being fitted. The posterior probability distributions are estimated using the affine invariant MCMC sampler implemented in \texttt{emcee} \citep{2010CAMCS...5...65G, 2013PASP..125..306F}, with 40 walkers evolved over 2000 steps for each clump. The first 1000 steps are discarded as burn-in. The median values for $v_0$ and the line FWHM (computed from $\sigma_v$) are listed in Table \ref{tab:clumpproperties}, with error bars based on the 16th and 84th percentiles of the samples. The median values and error bars for $I_\text{peak}$ are converted to brightness temperatures using the full Planck equation and also listed in Table \ref{tab:clumpproperties}. The spectra and best fits are plotted in Figure \ref{fig:clumpspectra}. 

\begin{figure*}
\begin{center}
\includegraphics{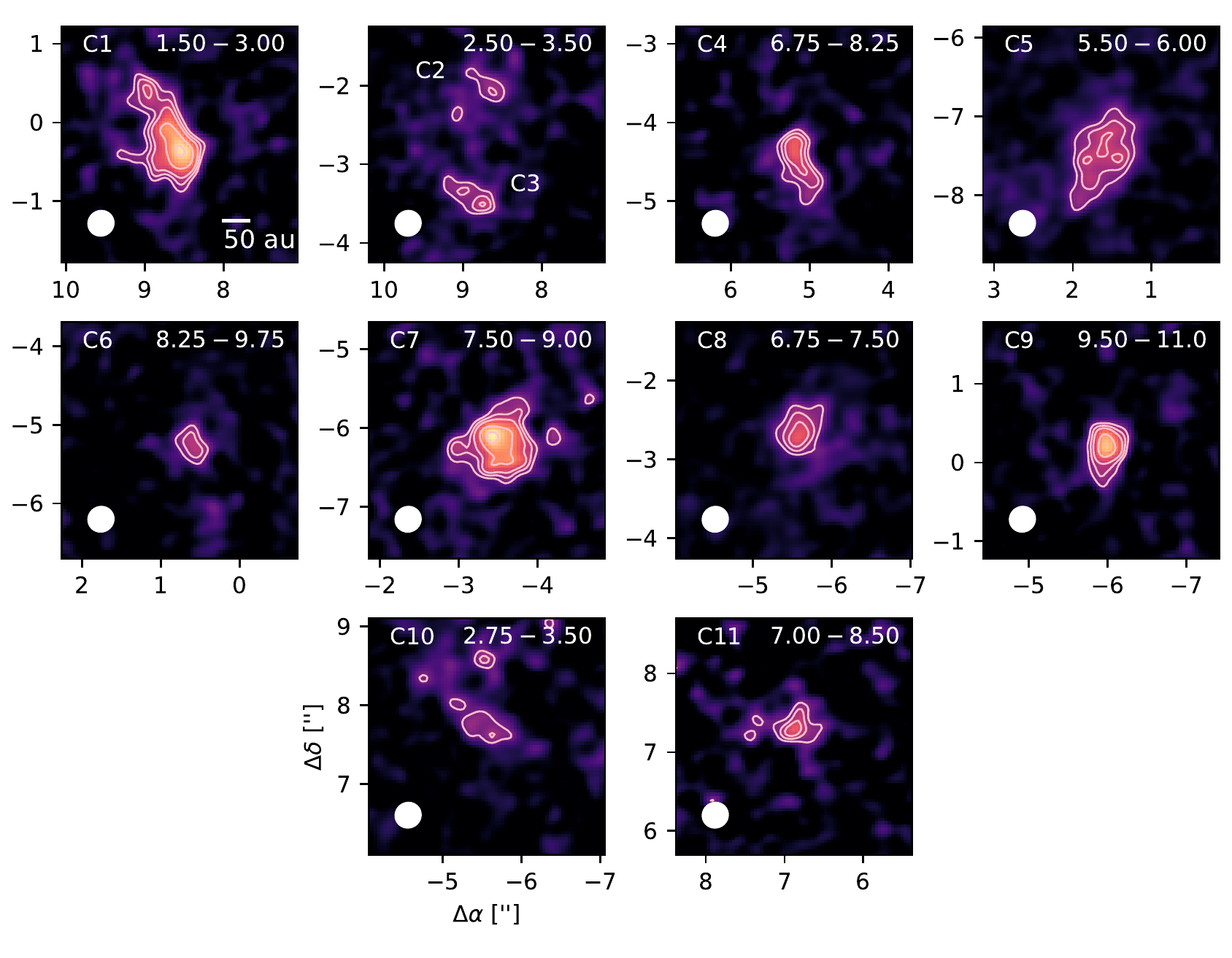}
\end{center}
\caption{Integrated intensity maps of the $^{12}$CO clumps observed around RU Lup. C2 and C3 are shown in the same panel due to their proximity. Contours are drawn at $[3, 4, 5, 7, 9]\sigma$, where $\sigma=2$ mJy beam$^{-1}$ km s$^{-1}$ is the typical noise level measured in the integrated intensity maps. The top right corner of each panel is labeled with the velocity integration range (in km s$^{-1}$). The axes are labeled with the offsets (in arcseconds) relative to the RU Lup disk continuum peak. The synthesized beam is drawn in the lower left corner of each panel. \label{fig:clumpgallery}}

\begin{center}
\includegraphics{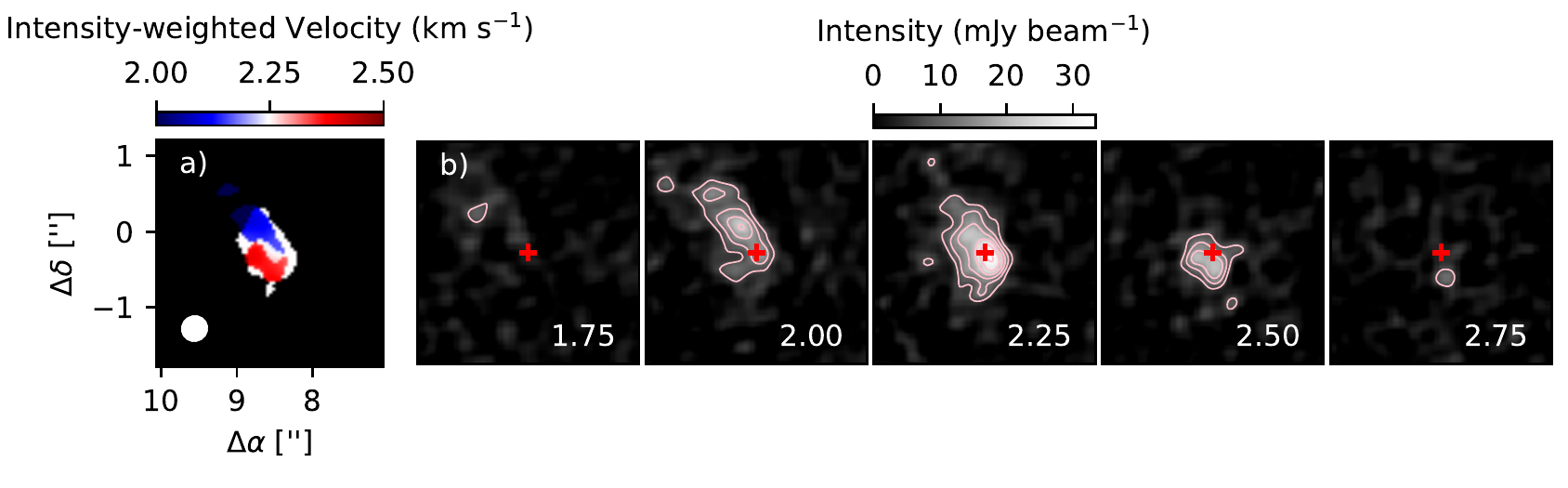}
\end{center}
\caption{Intensity-weighted velocity map (a) and $^{12}$CO channel maps (b) showing the velocity gradient across the C1 clump. Contours are drawn at [3, 5, 7, 9, 11]$\sigma$, where $\sigma=2.5$ mJy beam$^{-1}$ is the noise measured in the $^{12}$CO image cube. \label{fig:clumpgradient}}

\end{figure*}

To estimate the sizes and integrated fluxes of the clumps, we produce integrated intensity maps with velocity integration ranges set to 1.5 km s$^{-1}$ ($\sim2\times$ the FWHM of the widest line), roughly centered around the best-fit $v_0$. For C2, C3, C8, and C10, the velocity integration ranges are reduced to exclude cloud-contaminated channels. Since C5 lies entirely within the cloud-contaminated velocity range, its integrated intensity map only includes the channels where emission is detected above the $5\sigma$ level. Thus the emission measured for these clumps should be regarded as lower bounds. To avoid biasing these measurements, no clipping or masking is applied to channels within the integration range. The maps are shown in Figure \ref{fig:clumpgallery}. The integrated intensity maps are fit with two-dimensional Gaussians using the \texttt{imfit} task in CASA. This simple functional form is chosen to describe the clumps due to their low S/N. The FWHM (deconvolved from the beam) and the integrated flux computed from the best-fit Gaussians are reported in Table \ref{tab:clumpproperties}. The \texttt{imfit} task estimates uncertainties using the methods outlined in \citet{1997PASP..109..166C}. 

The clumps occur at a large range of LSRK velocities ($\sim2-10$ km s$^{-1}$), but most of them are redshifted relative to the systemic velocity ($\sim4.5$ km s$^{-1}$). There are no obvious trends between the clump velocities and their projected distance from RU Lup. The positions of some of the clumps may suggest a relationship with one another. Given the proximity and similar kinematics of C2 and C3 (and C1 to a lesser extent), they may be part of an underlying structure that is not detected at the sensitivity of the current observations. In the integrated intensity map, the positions of C9, C8, C7, C5, C4, and C1 appear to lie along an arc that connects with one of the branches of the S1 spiral arm. The kinematics, though, do not favor this interpretation. The spiral arms identified in Section \ref{sec:spiralarms}  are largely blueshifted relative to the systemic velocity, and the degree of blueshift tends to increase with distance from the star for a given arm. In contrast, the clump velocities do not change smoothly along their apparent ``arc.'' For example, C8 lies in between C7 and C9, but C8 appears at a smaller redshift than either C7 or C9. It is interesting to note, though, that the redshifted clumps tend to be southwest of RU Lup, consistent with the possible northeast-to-southwest spiral arm velocity gradient discussed in Section \ref{sec:spiralarms}. This kinematic behavior is reminiscent of the bipolar outflows from Class 0/I protostars, although bipolar protostellar outflows tend to be more collimated than the emission from RU Lup \citep[e.g.,][]{2014ApJ...783...29D}.   

To check whether the clumps are bound to RU Lup, we compute the escape velocity ($\sqrt{\frac{2GM_\ast}{r}}$) at a distance of 840 au (i.e., the projected distance of the clump closest to RU Lup) and compare it to the velocity offsets of the clumps relative to the systemic velocity ($\sim4.5$ km s$^{-1}$). Mass estimates for RU Lup have ranged from $0.2-0.5$ $M_\odot$ based on analyses of line kinematics \citep{2018AA...616A.100Y} and $0.6-1.2$ $M_\odot$ based on evolutionary models \citep[e.g.,][]{2017AA...600A..20A, 2018ApJ...865..157A}. Although dynamical mass estimates of T Tauri stars can often attain high precision \citep[e.g.,][]{2012ApJ...759..119R, 2015ApJ...806..154C}, the large uncertainty in RU Lup's stellar mass is likely in large part due to the non-Keplerian gas motion as well as its low inclination. Taking the 1.2 $M_\odot$ value to be conservative, we estimate that the escape velocity at 840 au is $\sim1.6$ km s$^{-1}$. This implies that C1, C4, C6, C7, C8, C9, and C11 are not bound to RU Lup. C2, C3, C5, and C10 appear to be bound to RU Lup if a stellar mass of 1.2 $M_\odot$ is adopted, but they would be unbound if a somewhat smaller stellar mass in the estimated range were adopted ($\sim0.8$ $M_\odot$). Higher precision stellar mass measurements will be necessary to determine whether these CO clumps are indeed bound to RU Lup. 

The clumps have similar peak brightness temperatures, with measurements ranging from $\sim8-13$ K. These measurements do not account for beam dilution, so the intrinsic brightness temperatures may be higher, but the measurements at least set a lower bound for the gas kinetic temperatures. It is plausible that the clumps are optically thick, since gas excitation temperature estimates from large-scale maps of the Lupus Molecular Cloud Complex range from $\sim8-10$ K \citep[e.g.,][]{2009ApJS..185...98T}. The clump brightness temperatures are also similar to those of the spiral arms. (Note that given the $5\sigma$ clump detection criterion, the measured peak brightness temperatures must be at least $\sim7$ K).

The linewidths of the clumps place an upper limit on the gas kinetic temperatures. If the CO line broadening is due entirely to thermal motion, then the maximum kinetic temperature is 

\begin{equation}
T^\text{max}_\text{kin} = \frac{m_\text{CO}}{(8\ln 2) k_B }\times (\text{FWHM})^2. 
\end{equation}


The measured linewidths (ranging from $\sim0.3-0.8$ km s$^{-1}$) translate to maximum kinetic temperatures of $50-390$ K, which are much larger than the measured brightness temperatures. These are very loose upper bounds, since other sources of line broadening could include saturation of the line core due to high optical depth, turbulence, rotation, and/or the presence of multiple unresolved velocity components \citep[e.g.,][]{1983ApJ...271..143D, 1986ApJ...303..356A}. A velocity gradient can most clearly be seen for C1, one of the largest and brightest clumps. The intensity-weighted velocity map and channel maps show that  the gas becomes more redshifted from northeast to southwest (Figure \ref{fig:clumpgradient}). The velocity signatures of the other clumps are generally more ambiguous due to their lower signal-to-noise ratios and smaller sizes. The S/N of the clumps does not allow the optical depth to be discerned readily on the basis of line shape, but as noted earlier, the line brightness temperatures are high enough to suggest that the clumps are optically thick. Given that the $^{12}$CO clumps are detected at the $5-10\sigma$ level and that the $^{12}$C/$^{13}$C ratio in the local ISM is $\sim69$ \citep[e.g.,][]{1999RPPh...62..143W}, the non-detection of $^{13}$CO clump emission does not provide strong constraints on the $^{12}$CO optical depth.

Most of the clumps appear to be spatially resolved, but are at most spanned by a few synthesized beams. Several of the clumps, such as C1, C4, and C5, are clearly elongated. C7 appears to be round, while the other clumps are too small to discern their shape. The size and shape of C10 are challenging to define, since it appears to have some diffuse components, but it is still counted as a clump because no emission connecting it to a larger structure is detected. 

\subsubsection{Column density and mass constraints}
The CO emission may be used to loosely estimate bounds on the masses of RU Lup's complex gas structures. We first consider the envelope-like structure, which is weakly detected in $^{13}$CO (see Figure \ref{fig:inner13CO}). Given that the $^{13}$CO emission is faint and that the envelope-like structure is not visible in C$^{18}$O, we assume that $^{13}$CO $J=2-1$ is optically thin in this region. The gas-phase column density of a molecular species in local thermodynamic equilibrium can be estimated from an optically thin transition with

\begin{equation}\label{eq:coldensity}
N_\text{tot, thin} = \frac{4\pi}{A_{ul} hc}\times\frac{Q(T_\text{rot})e^{\frac{E_u}{T_\text{rot}}}}{g_u}\times \frac{\int S_\nu(v) dv}{\Omega} ,
\end{equation}
where $A_{ul}$ is the Einstein $A$ coefficient for the transition of interest, $g_u$ is the upper state degeneracy, $Q(T_\text{rot})$ is the partition function evaluated at the rotational temperature $T_\text{rot}$, $E_u$ is the upper state energy (in K), $\int S_\nu(v) dv$ is the velocity-integrated flux density, and $\Omega$ is the solid angle of the emitting area \citep[e.g.,][]{1999ApJ...517..209G, 2008AA...488..959B}. The molecular constants $A_{ul}$, $E_u$, and $g_u$ are listed in Table \ref{tab:constants}.

\begin{deluxetable}{ccccccc}
\tablecaption{Molecular constants\tablenotemark{a}\label{tab:constants}}
\tablehead{
\colhead{Species} & \colhead{Transition}& \colhead{Frequency}&\colhead{$A_{ul}$}&\colhead{$g_u$}&\colhead{$E_u$}\\
&&\colhead{(GHz)}&\colhead{(s$^{-1}$)}&&\colhead{(K)}}
\startdata
$^{13}$CO & $J=2-1$ &220.3986842&$6.075\times10^{-7}$ & 10 & 15.9\\
$^{12}$CO & $J=2-1$ &230.5380000& $6.911\times10^{-7}$ & 5& 16.6\\
\enddata
\tablenotetext{a}{Compiled from the Cologne Database for Molecular Spectroscopy \citep{2001AA...370L..49M, 2005JMoSt.742..215M}, based on data from \citet{1994ApJS...95..535G}, \citet{1997JMoSp.184..468W}, and \citet{2000JMoSp.201..124K}.}
\end{deluxetable}

Since the envelope-like structure overlaps with both disk and ambient cloud emission, it is not straightforward to estimate the column density simply by measuring an integrated $^{13}$CO flux and applying Equation \ref{eq:coldensity}. Instead, we estimate the column density that would be necessary for a $\sim5\sigma$ (10 mJy beam$^{-1}$) detection of $^{13}$CO $J=2-1$ in individual channels, as seen for portions of the envelope-like structure in Figure \ref{fig:inner13CO}. Assuming that $^{12}$CO is optically thick and that therefore its brightness temperature traces the gas temperature (see Figure \ref{fig:spiralsannotated}), we take $T_\text{rot}$ to be $\sim35$ K. We compute $\frac{\int S_\nu(v) dv}{\Omega}$ assuming a Gaussian line profile with a peak intensity of 10 mJy per beam (i.e., $\Omega$ is the solid angle of the synthesized beam). Since the linewidth (FWHM) of the envelope-like structure cannot be measured well in the presence of cloud contamination and disk emission, we assume that the linewidth is 0.2 km s$^{-1}$, comparable to what has been measured in some protoplanetary disks \citep[e.g.,][]{2016AA...592A..49T}. Uncertainty in the linewidth contributes to uncertainty in column density estimates by a factor of a few, so the column density is at best accurate within an order-of magnitude. The partition function, which is calculated by interpolating the tabulated values in the Cologne Database for Molecular Spectroscopy, is $Q(35) = 27.2$ \citep{2001AA...370L..49M, 2005JMoSt.742..215M}. The resulting $^{13}$CO column density is estimated to be $\sim3\times10^{14}$ cm$^{-2}$. Assuming an ISM-like ratio of $^{12}$C:$^{13}$C of 69 and $^{12}$CO:H$_2$ of $10^{-4}$, the $^{13}$CO column density estimate corresponds to an H$_2$ column density of $\sim2\times10^{20}$ cm$^{-2}$. Taking this to be the average H$_2$ column density over a circular area with a radius of $260$ au and assuming that the envelope-like structure is predominantly H$_2$, we estimate the overall mass of the envelope-like structure to be $\sim2\times10^{-5}$ $M_\odot$.

As noted in Section \ref{sec:innerregions}, the kinematics of the envelope-like structure are complex, and it is not clear whether material from it is infalling. However, we can estimate the rate at which mass could be transferred to the disk if the envelope were freely falling. The free-fall velocity is $v_\text{ff}(r) = \sqrt{\frac{2GM_\ast}{r}}$. Under the assumption of spherical symmetry, the mass infall rate at a distance $r$ is
\begin{equation}
\dot M_\text{infall}(r) = 4\pi \mu m_H r^2 n_\text{gas}(r) v_\text{ff}(r),
\end{equation}
where $\mu m_H$ is the mean molecular weight \citep[e.g.,][]{2013AA...558A.126M}. We take $\mu$ to be 2.37. The volume of the envelope-like structure is approximated as that of a spherical shell with an inner radius of 120 au (corresponding to the outer edge of the C$^{18}$O emission tracing the Keplerian disk) and an outer radius of 260 au. A mass of $\sim2\times10^{-5}$ $M_\odot$ corresponds to an average gas number density of $\sim5\times10^4$ cm$^{-3}$. Assuming a stellar mass between 0.6 and 1.2 $M_\odot$ \citep{2014AA...561A...2A, 2018ApJ...865..157A}, the free-fall velocity at 120 au is $\sim3-4$ km s$^{-1}$. Thus the infall rate onto the disk at $r=120$ au would be $\dot M_\text{infall}\sim3\times10^{-8}-5\times10^{-8}$ $M_\odot$ yr$^{-1}$. For reference, RU Lup's stellar accretion rate is estimated to range from $4.0\times10^{-8}$ to $1.1\times10^{-7}$ $M_\odot$ yr$^{-1}$, depending on the model \citep{2017AA...600A..20A}.

A number of caveats apply to the estimates in this section. In general, mass and column density estimates involving cloud-contaminated emission should be treated with caution. \citet{2016ApJ...828...46A} and \citet{2017AA...599A.113M} have suggested that a significant fraction of the RU Lup disk's warm gas-phase CO is chemically depleted, based on order-of-magnitude discrepancies in disk mass estimates derived from CO isotopologue fluxes and sub-millimeter continuum emission. Since continuum emission is not detected at the radii where the envelope-like structure appears in molecular emission, it is not clear whether CO is also depleted in this region. Nevertheless, if the assumed $^{12}$CO:H$_2$ ratio is indeed too high, then the mass of the envelope-like structure is underestimated. We have also not accounted for radial density and temperature gradients. However, we can check whether the values are broadly reasonable. Given that the optically thinner CO isotopologues are dominated by emission from the Keplerian disk, the mass of the envelope-like structure should be much smaller than that of the disk. Based on the dust mass estimated from millimeter continuum emission and assuming a 100:1 gas-to-dust ratio, the total mass of the RU Lup disk is 0.03 $M_\odot$ \citep{2016ApJ...828...46A}, indeed several orders of magnitude higher than the estimated mass of the envelope-like structure. We can also check whether the estimated gas column density of the envelope-like structure is reasonable given the low amounts of material inferred for the line-of-sight to RU Lup. The \ion{H}{1} log column density towards RU Lup is estimated from Lyman-$\alpha$ absorption to be $\log \text{N(H I)}=20.0\pm0.15$, where N(\ion{H}{1}) is in cm$^{-2}$ \citep{2005AJ....129.2777H}. If we assume that the \ion{H}{1} originates entirely from the envelope-like structure, then the H$_2$ column density we estimated for the envelope-like structure can be accounted for by gas that is at least 2/3 molecular hydrogen by number. Given the bright CO emission detected in this region, we expect the envelope-like structure to be overwhelmingly molecular rather than atomic hydrogen. Thus, the H$_2$ column density estimated for the envelope-like structure does not appear to be excessive. (\citealt{2005AJ....129.2777H} also directly derive an H$_2$ column density from UV spectroscopy, but this is largely sensitive to gas between 100 and 300 K and therefore would not trace the bulk of the envelope-like structure.) The Lyman-$\alpha$ absorption measurements, however, do not rule out the possibility that the mass of the envelope-like structure could be significantly higher, since the line-of-sight to the star is not necessarily representative of the overall emitting region of the envelope-like structure.

Meanwhile, the clumps and spiral arms are only detected in $^{12}$CO emission, which is likely optically thick. Thus, $^{12}$CO can only be used to estimate a lower bound on the column densities. The column density of a transition with an optical depth of $\tau$ can be expressed as 
\begin{equation}
N_\text{tot}=N_\text{tot, thin}\times \frac{\tau}{1-e^{-\tau}},
\end{equation}
where $N_\text{tot,thin}$ is given by Equation \ref{eq:coldensity} \citep{1999ApJ...517..209G}.

The values of $A_{ul}$, $E_u$, and $g_u$ for $^{12}$CO $J=2-1$ are listed in Table \ref{tab:constants}. Selecting values representative of the clumps, we set $\Omega$ to the solid angle corresponding to a circle with a radius of 50 au and $\int S_\nu(v) dv$ to 50 mJy km s$^{-1}$. $T_\text{rot}$ is taken to be 10 K, the typical $^{12}$CO brightness temperature of a clump. The partition function, which is calculated by interpolating the tabulated values in the Cologne Database for Molecular Spectroscopy, is $Q(10) = 3.97$ \citep{2001AA...370L..49M, 2005JMoSt.742..215M}. Assuming a lower bound of $\tau=3$, the clump-averaged CO column density lower bound would be $N_\text{tot}\sim 8\times 10^{15}$ cm$^{-2}$. An ISM-like $^{12}$CO:H$_2$ abundance ratio of $10^{-4}$ yields a clump-averaged H$_2$ column density lower bound of $\sim 8\times 10^{19}$ cm$^{-2}$, assuming that the clumps are predominantly molecular hydrogen. For a representative clump radius of 50 au, this H$_2$ column density corresponds to a clump mass lower bound of $5\times10^{26}$ g, or $\sim0.1$ $M_\earth$.

Beyond optical depth uncertainties, the clump masses could also be orders of magnitude larger than this lower bound since freezeout and chemical depletion may decrease the $^{12}$CO:H$_2$ abundance. The assumed gas temperature of 10 K is well below the CO freezeout temperature ranges of $\sim20-25$ K estimated for protoplanetary disks \citep[e.g.,][]{2013Sci...341..630Q, 2015ApJ...813..128Q, 2017NatAs...1E.130Z}, so the $^{12}$CO:H$_2$ ratio could be far less than $10^{-4}$ if the bulk of the CO in the clumps is frozen out. On the other hand, since CO depletion timescales can be tens of thousands of years \citep[e.g.,][]{1995ApJ...441..222B}, CO gas can still be abundant if the clumps formed recently, if the clumps formed at warmer temperatures closer to the star and were transported outward, or if non-thermal processes such as UV photo-desorption are active. As noted earlier, CO isotopologue analysis by \citet{2016ApJ...828...46A} and \citet{2017AA...599A.113M} suggest that CO is depleted by an order of magnitude in the warm gas of the RU Lup disk. That said, those observations were only sensitive to distances up to $\sim100$ au from the star, so the disk and clumps do not necessarily have the same chemical compositions if they are subject to very different physical conditions. A similar analysis of the gas column density lower bounds is applicable to the spiral arms, which have $^{12}$CO brightness temperatures comparable to the clumps.

The non-detection of millimeter continuum emission can be used to estimate an upper limit on the gas surface density in the spirals and clumps, if we assume that the dust is optically thin and that the gas-to-dust ratio is interstellar (100). The intensity of optically thin dust emission is approximately $I_\nu \approx \kappa_\text{abs}\Sigma_d B_\nu(T_d)$, where $\kappa_\text{abs}$ is the dust opacity, $\Sigma_d$ is the dust surface density, and $T_d$ is the dust temperature. For a given intensity value, upper limits for $\Sigma_d$ can be calculated by using the minimum plausible opacity and temperature values. We adopt the DSHARP dust opacities from \citet{2018ApJ...869L..45B} and assume a standard power-law size distribution of the form $n(a)\propto a^{-q}$ between some minimum grain size $a_\text{min}$ and maximum size $a_\text{max}$. At millimeter wavelengths, the dust opacity as a function of $a_\text{max}$ peaks near 1 mm. We assume that a population of millimeter-sized grains or larger is unlikely to be present in the clumps and spiral arms, which occur at large distances from RU Lup. Larger dust particles are expected to be found closer to the star because dust coagulation is more favorable in higher-density regions and larger particles tend to drift inward due to decoupling from the gas \citep[e.g.,][]{2008AA...480..859B}. For $a_\text{max}$ smaller than $\sim100$ $\mu$m, the dust opacity as a function of $a_\text{max}$ is nearly flat and does not have a strong dependence on $q$. Thus, we take our lower bound on $\kappa_\text{abs}$ at $\lambda=1.3$ mm to be 0.4 cm$^2$ g$^{-1}$, which is the opacity value corresponding to $a_\text{max}=100$ $\mu$m, $a_\text{min}=0.1$ $\mu$m, and $q=3.5$ (i.e., the interstellar size distribution power law derived by \citet{1977ApJ...217..425M}). The lower bound on $T_d$ is set to 10 K, the typical $^{12}$CO brightness temperature of the clumps and spiral arms. The 3$\sigma$ upper limit for the clump intensity is estimated by measuring the fluxes at random off-source positions in the continuum image inside circles with a radius of 50 au, multiplying the standard deviation (0.04 mJy) by 3, and dividing by the area of the aperture to obtain an average intensity. With the assumed temperature and dust opacity, we obtain a $\Sigma_d$ upper limit of 0.005 g cm$^{-2}$. For an interstellar gas-to-dust ratio of 100, this corresponds to a gas surface density upper limit of 0.5 g cm$^{-2}$. Thus, a typical clump (radius of $\sim50$ au) should be less than 150 M$_\earth$, or $\sim0.5$ $M_\text{Jup}$. (If we adopt assumptions typical for protoplanetary disks with grain growth up to millimeter sizes, such as $T_d=20$ K and $\kappa_\text{1.3 mm}=2.3$ cm$^{2}$ g$^{-1}$ \citep[e.g.,][]{2016ApJ...828...46A}, the mass upper limit decreases by an order of magnitude). If the gas is predominantly molecular hydrogen, a gas surface density of 0.5 g cm$^{-2}$ translates to a molecular hydrogen column density of $\sim10^{23}$ cm$^{-2}$, or at most a CO column density of $\sim 10^{19}$ cm$^{-2}$ if we assume a $^{12}$CO:H$_2$ ratio of 10$^{-4}$. This is several orders of magnitude higher than the $^{12}$CO column density lower limit estimated from the line fluxes, demonstrating that 1) the constraints on the clump masses are quite loose, and 2) the non-detection of continuum emission at the CO clumps and spiral arms is still consistent with standard $^{12}$CO:H$_2$ and dust-to-gas ratios. 

Under the assumption that the clumps are discrete, one can in principle assess whether they are internally gravitationally bound by computing the virial parameter:
\begin{equation}
\alpha = \frac{5\sigma^2 R}{GM},
\end{equation}
where $\sigma$ is the one-dimensional velocity dispersion of the clump (such that $\sigma$ represents the standard deviation of a Gaussian distribution), $R$ is the clump radius, and $M$ is the clump mass \citep[e.g.,][]{1992ApJ...395..140B}. If $\alpha >1$, then the clump is not internally gravitationally bound. Following \citet{1992ApJ...384..523F}, we compute the velocity dispersion with

\begin{equation}
\sigma^2 = \frac{\text{FWHM}^2_\text{CO}}{8\ln 2} + kT\left(\frac{1}{\mu m_H}-\frac{1}{m_\text{CO}}\right). 
\end{equation}
Setting FWHM$_\text{CO}=0.5$ km s$^{-1}$, $T=10$ K, $R=50$ au, and $M=0.1$ $M_\earth$ as representative clump values, we obtain $\alpha\sim7\times10^4$. Allowing for the possibility that the clump masses are underestimated by a few orders of magnitude due to the $^{12}$CO optical depth and the assumed $^{12}$CO:H$_2$ ratio, $\alpha\sim100$. At face value, this suggests that the clumps are not internally gravitationally bound and will disperse quickly. That said, the velocity dispersion is also highly uncertain due to potential broadening of the $^{12}$CO linewidth by optical depth effects and/or rotation, which will also bias $\alpha$ values upward. Higher-quality observations will be necessary to draw firmer conclusions about the clump properties.

\begin{figure*}
\begin{center}
\includegraphics{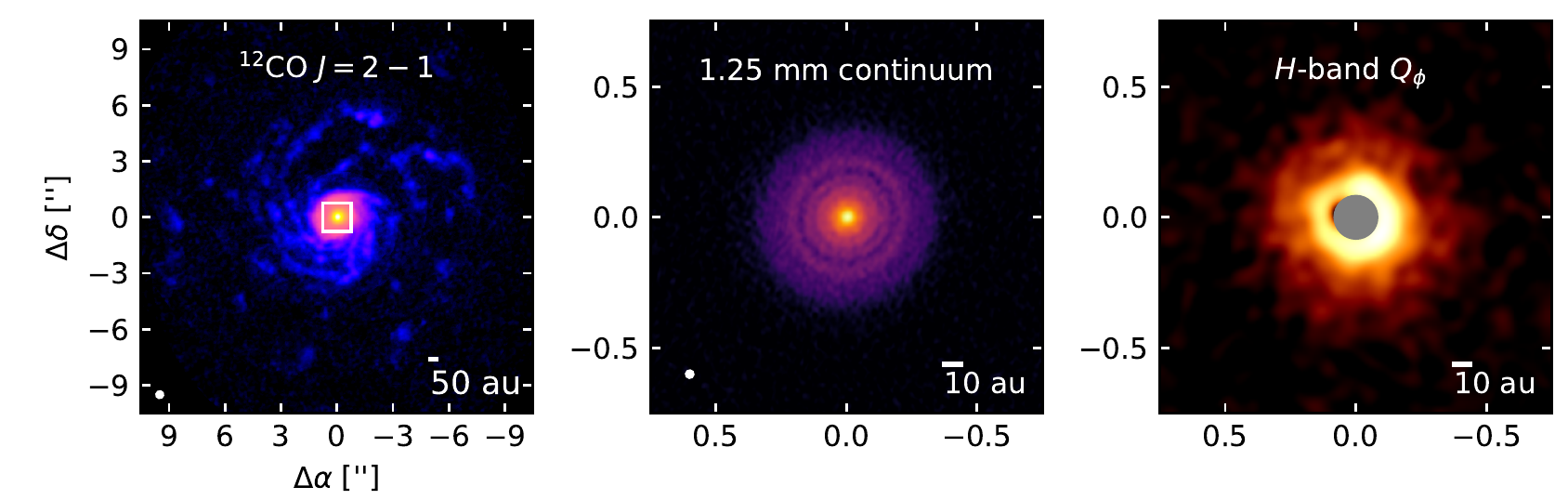}
\end{center}
\caption{A comparison of $^{12}$CO observations toward RU Lup with millimeter continuum and scattered light observations. \textit{Left}: $^{12}$CO map from this work. The white box shows the size of the other two panels. \textit{Middle}: High-resolution millimeter continuum image of the RU Lup disk from DSHARP \citep{2018ApJ...869L..41A}. \textit{Right}: $H$-band $Q_\phi$ image from DARTTS-S \citep{2018ApJ...863...44A}. An asinh stretch is used for all the images to make the faint structure visible.  \label{fig:dustcomparison}}
\end{figure*}

\subsection{Comparison of $^{12}$CO and dust observations\label{sec:RULupdustcomparison}}
RU Lup's $^{12}$CO emission morphology differs dramatically from the appearance of the system in millimeter continuum (predominantly tracing millimeter-sized dust grains) or in scattered light (predominantly tracing sub-micron-sized dust grains). Figure \ref{fig:dustcomparison} compares the $^{12}$CO emission to the high-resolution millimeter continuum image from the DSHARP survey \citep{2018ApJ...869L..41A} and the VLT-SPHERE  $H$-band $Q_\phi$ image from the DARTTS-S survey \footnote{Based on observations collected at the European Southern Observatory under ESO programme  096.C-0523(A).} \citep{2018ApJ...863...44A}. The $Q_\phi$ image is the version convolved with a Gaussian with a FWHM of 75 mas, as shown in Figure 2 of \citet{2018ApJ...863...44A}.

The millimeter continuum is symmetric, exhibiting multiple annular gaps and rings when observed at a spatial resolution of $\sim4$ au. The millimeter continuum disk extends to a radius of $\sim63$ au \citep{2018ApJ...869L..42H}, lying entirely within the Keplerian disk traced by C$^{18}$O. In other words, no continuum emission is detected in the regions where $^{12}$CO emission traces the ``envelope-like'' structure, the spiral arms, or the clumps. Likewise, the scattered light observations from \citet{2018ApJ...863...44A} only show signal up to $\sim75$ au from the star, and exhibit no evidence for substructures. In fact,  \citet{2018ApJ...863...44A} refer to the RU Lup disk as the ``most unremarkable'' in their survey. They interpret the apparent brightness enhancement on the southwest side of the disk as evidence that this side is closer to the observer, with the caveat that the signal-to-noise ratio is not high enough to be definitive. The CO isotopologue observations unfortunately do not clarify whether their inferred orientation is correct, since RU Lup's inclination is very low and the Keplerian disk is not spatially well-resolved.

\begin{figure}
\begin{center}
\includegraphics{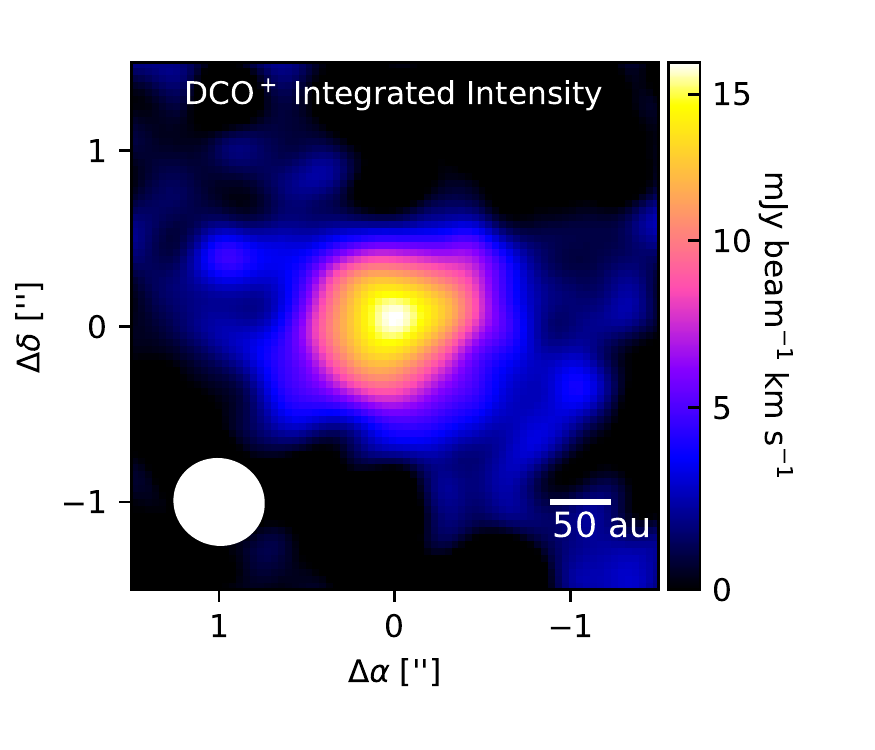}
\end{center}
\caption{Integrated intensity map of DCO$^+$ $J=3-2$. The synthesized beam is plotted in the lower left corner. Offsets from the phase center are marked on the axes. \label{fig:DCOpmommap}}
\end{figure}

\begin{figure*}
\begin{center}
\includegraphics{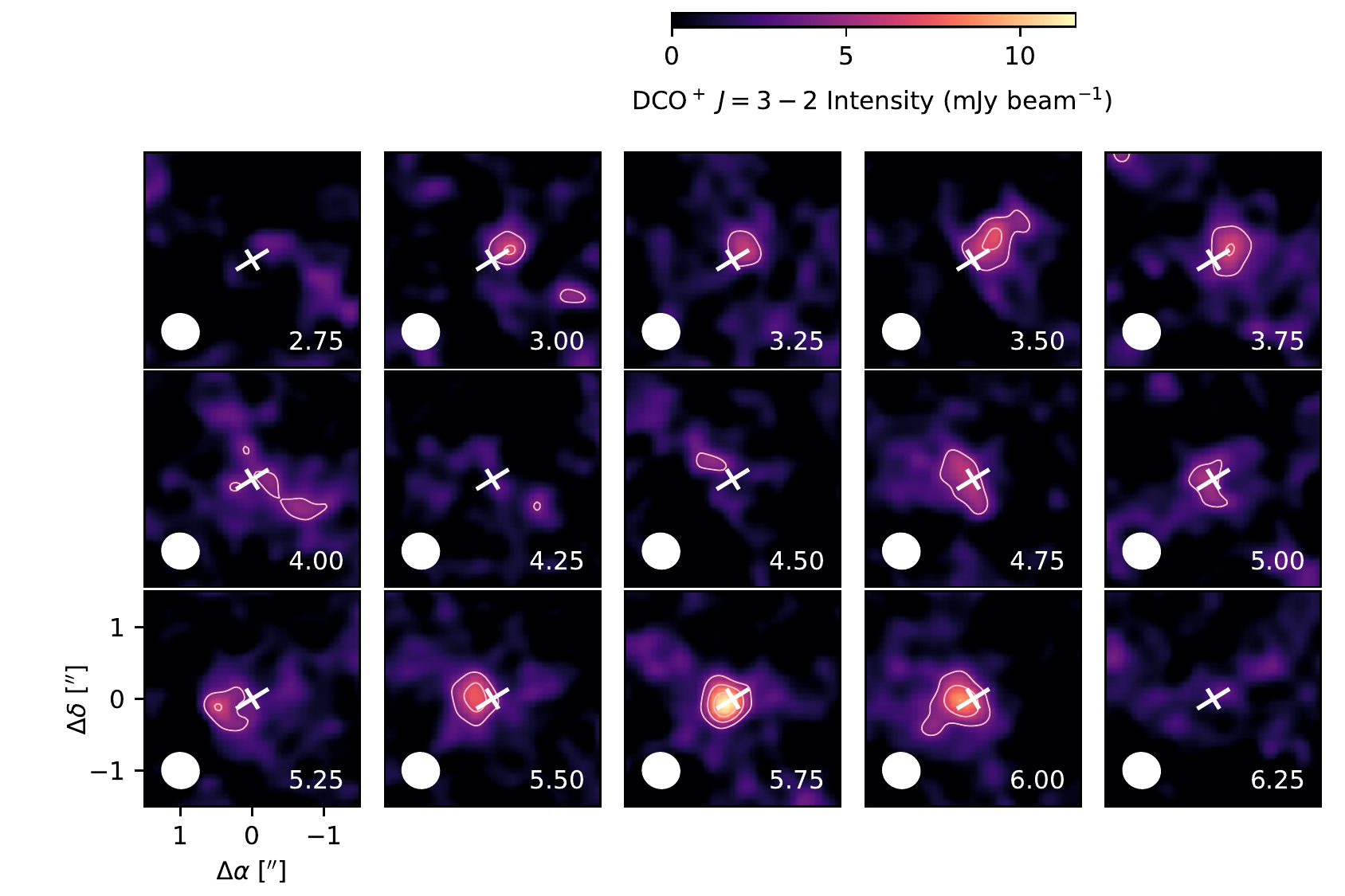}
\end{center}
\caption{Channel maps of DCO$^+$ $J=3-2$ emission. The white cross marks the position of the continuum peak, with the longer line denoting the disk position angle. The synthesized beam is plotted in the lower left corner of each panel. Offsets from the phase center are marked on the axes. Contours are drawn at the [3, 5, 7]$\sigma$ levels. \label{fig:DCOpchanmaps}}
\end{figure*}

\subsection{DCO$^+$ emission}

The DCO$^+$ integrated intensity map (Figure \ref{fig:DCOpmommap}) is produced in a manner similar to the $^{13}$CO and C$^{18}$O maps, but only includes channels from 2.75 to 6.25 km s$^{-1}$ because no disk emission is detected at the $3\sigma$ level outside this range. The flux is also measured in a similar manner and reported in Table \ref{tab:imageproperties}. The S/N of the DCO$^+$ emission is not high enough to generate a useful intensity-weighted velocity map, but the DCO$^+$ channel maps are shown in Figure \ref{fig:DCOpchanmaps}. The maps illustrate that emission northwest of the disk center is blueshifted relative to the systemic velocity and emission southeast of the disk center is redshifted, consistent with the Keplerian rotation pattern established by C$^{18}$O. 

Like C$^{18}$O, the DCO$^+$ emission is compact and does not trace the spiral arms or clumps. In contrast to DCO$^+$ emission observed toward most other disks at comparable spatial resolution \citep[e.g.,][]{2013AA...557A.132M, 2017ApJ...835..231H, 2017ApJ...838...97M}, RU Lup's DCO$^+$ emission does not exhibit any obvious substructure. The formation of DCO$^+$ is disfavored at higher temperatures \citep[e.g.,][]{2000AA...361..388R, 1984ApJ...287L..47D}, which is expected to result in a central cavity in DCO$^+$ emission because the disk temperature generally decreases with distance from the star \citep[e.g.,][]{2013AA...557A.132M}. In this framework, the absence of resolved substructure in the RU Lup disk would imply that it is colder compared to disks with resolved DCO$^+$ substructure. However, its highly luminous SED suggests that the RU Lup disk is much warmer than the typical T Tauri disk  \citep{2018ApJ...869L..41A}. Alternatively, the centrally peaked DCO$^+$ emission may trace contributions from formation pathways active at warm temperatures, such as reactions between CO and hydrocarbon cations \citep[e.g.,][]{2015ApJ...802L..23F}. Observing additional transitions of DCO$^+$ would be useful for measuring the excitation temperature and establishing whether RU Lup's thermal structure or DCO$^+$ chemistry differs from other disks. 

Scaling for distance, RU Lup's DCO$^+$ flux is $3-10$ times weaker than DCO$^+$ $J=3-2$ fluxes measured in other disks \citep[e.g.,][]{2010ApJ...720..480O, 2011ApJ...734...98O, 2017ApJ...835..231H, 2018AA...614A.106C}. This is at least partially a selection effect, since earlier millimeter interferometers such as the Submillimeter Array did not place strong upper limits on DCO$^+$ non-detections \citep[e.g.,][]{2010ApJ...720..480O, 2011ApJ...734...98O}, and subsequent ALMA observations of DCO$^+$ in disks have tended to focus on systems that were already known to have bright DCO$^+$ emission and/or are thought to have massive disks \citep[e.g.,][]{2017ApJ...835..231H, 2017AA...606A.125S}. The millimeter continuum luminosity of RU Lup is comparable to other disks with bright DCO$^+$ emission \citep[e.g.,][]{2018ApJ...865..157A}, so it is unlikely that RU Lup's weak DCO$^+$ emission is due to a substantially lower disk mass. Another explanation could be that RU Lup's disk is more weakly ionized, which can be investigated by measuring the abundances of other molecular ions \citep[e.g.,][]{2011ApJ...743..152O, 2015ApJ...799..204C}.

\section{Discussion \label{sec:discussion}}
\subsection{Origin of RU Lup's features}
The $^{12}$CO emission of RU Lup displays three unusual features: a non-Keplerian ``envelope-like'' structure surrounding the Keplerian disk, blueshifted spiral arms, and clumps located hundreds of au from the star. It is ambiguous whether these features have a common origin. Some of the explanations discussed below may explain multiple features, while others may only explain individual features. 
\subsubsection{Gravitational instability\label{sec:GI}}
Gravitational instability (GI) has been proposed to trigger spiral arms in massive disks, which can subsequently fragment and form clumps \citep[e.g.,][]{1997Sci...276.1836B, 2001ApJ...553..174G}. Clumps can be ejected at speeds exceeding the escape velocity \citep[e.g.,][]{2016AA...590A.115V}, perhaps accounting for the high clump speeds observed for RU Lup. 

While the Toomre $Q$ parameter \citep{1964ApJ...139.1217T} is typically used to evaluate whether a disk is gravitationally stable, doing so is not straightforward for RU Lup. In addition to the mass and temperature uncertainties noted earlier, it is unclear what the appropriate epicyclic frequency to adopt is (for Keplerian disks, the epicyclic frequency is simply the angular rotation frequency). However, RU Lup has other characteristics that, when considered together, make GI an intriguing explanation for its spiral arms and clumps.  Disks become more susceptible to GI at lower temperatures. As noted in Section \ref{sec:spiralarms}, the inner tips of the spiral arms connect to a clumpy ring of emission with a radius of $\sim260$ au. This morphology may indicate that the clumpy emission ring marks a transition region where spiral arms can form more readily, perhaps because temperatures have decreased enough for the system to become gravitationally unstable. Furthermore, GI is favored in more massive systems. While absolute disk mass measurements are a highly contested subject, RU Lup's bright millimeter continuum suggests that its disk ranks among the most massive in Lupus. Only two other spiral-armed disks have been detected in Lupus, albeit in millimeter continuum emission rather than $^{12}$CO: the IM Lup disk, which also ranks among the brightest disks in CO and continuum emission \citep{2018ApJ...869L..43H}, and the HT Lup disk, which has a nearby companion that is likely triggering its spirals \citep{2018ApJ...869L..44K}. GI is also expected to enhance stellar accretion rates \citep[e.g.,][]{2015ApJ...812L..32D, 2015ApJ...805..115V}. Out of a survey of 90\% of the Class II and transitional young stellar objects in Lupus, RU Lup exhibits the highest stellar accretion rate, ranging from $4.0\times10^{-8}$ to $1.1\times10^{-7}$ $M_\odot$ yr$^{-1}$ depending on the model \citep{2017AA...600A..20A}. Finally, based on observations dating back more than a century, RU Lup is notable for its irregular optical and infrared photometric variations spanning several magnitudes \citep[e.g.,][]{1916HarCi.196....1M, 1945ApJ...102..168J, 1974AA....33..399G, 1991AAS...87...89G}. \citet{2005ApJ...633L.137V} proposed that the large, irregular outbursts of FU Ori objects are due to the formation and infall of clumps in gravitationally unstable protostellar disks. A scaled-down version of this mechanism may be responsible for RU Lup's photometric irregularities. 

It is not evident, though, that the blueshifted spiral kinematic pattern can be explained by a GI origin; simulations by \citet{2015ApJ...812L..32D} indicate that the spiral arms should have approximately Keplerian kinematics. Furthermore, elevated accretion rate and luminosity variations are not uniquely signatures of GI, as detailed later in this section.

\subsubsection{Accretion from an envelope or cloud}
Spiral arms may be triggered through infall from a protostellar envelope or capture from molecular cloud material \citep[e.g.][]{2015ApJ...805...15B, 2015AA...582L...9L, 2019AA...628A..20D}. The infalling motion would manifest as non-Keplerian emission around the disk, and infall or capture of material onto the disk is expected to enhance stellar accretion rates \citep[e.g.,][]{2008AJ....135.2380T}. 

Given that infall, clumpy outflows, and arm-like structures are often observed in interferometric maps of molecular emission toward embedded Class 0/I sources \citep[e.g.,][]{2007ApJ...659..479J, 2011ApJ...740...45T, 2014ApJ...793....1Y, 2015Natur.527...70P}, we examine the question of whether RU Lup has been appropriately categorized as a more evolved Class II source. Traditionally, the evolutionary stage of a young star is determined by measuring its infrared SED slope $a = \sfrac{d\log \lambda F_\lambda}{d\log \lambda}$ \citep{1987IAUS..115....1L, 1994ApJ...434..614G}. Younger sources have positive slopes due to emission from their massive protostellar envelopes, and the slope decreases as the envelope dissipates. Based on the RU Lup SED collated by \citet{2018ApJ...869L..41A}, $a=-0.32\pm0.03$ between 2.2 and 22.1 $\mu$m, which falls on the border between a Flat Spectrum and Class II source according to the definitions from \citet{1994ApJ...434..614G}. Other than the high degree of variability, RU Lup's stellar spectrum is similar to other classical T Tauri stars \citep[e.g.,][]{2008hsf2.book..295C}. There is no obvious sign of a protostellar envelope surrounding RU Lup's spiral structures, although detecting one with ALMA in $^{12}$CO emission would be complicated by both the cloud contamination and spatial filtering. Nevertheless, it is notable that the Keplerian disk clearly dominates RU Lup's $^{13}$CO and C$^{18}$O emission, while outflows and infall often dominate in interferometric observations of Class I sources in these isotopologues and sometimes even in the optically thinner C$^{17}$O \citep[e.g.,][]{2019AA...626A..71A, 2020ApJ...891L..17Z}. \citet{2005AJ....129.2777H} estimate a very low extinction value of $A_v\sim0.07$, suggesting that there is little material along the line of sight. Even if RU Lup is surrounded by an envelope, it must be more tenuous than the envelopes of sources traditionally categorized as Class I. Thus, RU Lup appears to be more evolved than typical Class I sources, but may be at the tail end of envelope dispersal. 

Given that RU Lup's SED appears to be relatively evolved, it is intriguing to consider whether its spiral arms result from an encounter with a cloud fragment. Continued external delivery of material to protoplanetary disks has previously been explored as an explanation for accretion rate anomalies observed in pre-main sequence stars \citep[e.g.,][]{2005ApJ...622L..61P, 2014AA...566L...3S}, and more recently has received attention as a possible resolution to the apparent discrepancy between protoplanetary disk and exoplanetary system masses \citep[e.g.,][]{2018AA...618L...3M, 2019AA...628A..20D, 2020AA...633A...3K}.  The geometry of RU Lup's irregular, large-scale spiral arms bears some resemblance to the simulations shown in \citet{ 2019AA...628A..20D} and \citet{2020AA...633A...3K}. The velocity profile snapshots from \citet{2020AA...633A...3K}, though, show that most of the externally delivered gas is moving at or below Keplerian speeds, while at least some of the gas surrounding RU Lup exceeds the escape velocities. 

\subsubsection{Winds}
Based on the non-Keplerian shape of its $^{12}$CO 4.67 $\mu$m fundamental ($\nu=1-0$) ro-vibrational band spectrum and anomalously strong X-ray absorption, RU Lup has been proposed to be the site of a stellar or disk wind \citep[e.g.,][]{2007AA...473..229R, 2011ApJ...733...84P}. CO ro-vibrational emission originates largely from gas in the surface layers of the inner few au of the disk. The non-Keplerian ALMA $^{12}$CO emission, which traces the cool outer disk regions, may be an extension of the proposed inner disk wind.  

Clumps and spirals (in addition to rings) have been shown to form in simulations of disks with MHD winds \citep[e.g.,][]{2000ApJ...528..462H, 2011ApJ...735..122F, 2018MNRAS.477.1239S}. However, the spirals formed in MHD simulations are generally tightly-wrapped, unlike the wide spirals observed toward RU Lup. Furthermore, the clumps seen in MHD simulations are formed within the disk, not separated from the Keplerian disk like those seen toward RU Lup. These simulations are often limited to disk regions within 100 au of the star, which is much more compact than the extent of the gas around RU Lup. Expanding the outer boundaries of the simulations will be necessary to make more detailed comparisons to RU Lup. 

UV radiation, either from the central star or from the interstellar radiation field, can drive photoevaporative winds that remove gas from the surface layers of disks \citep[e.g.,][]{1994ApJ...428..654H, 2004ApJ...611..360A}. Photoevaporative winds have not to our knowledge been shown to trigger the formation of clumps or spiral arms in disks. One challenge with attributing the complex CO kinematics to a photoevaporative wind is that CO is often expected to be photodissociated in most of the gas expelled from the disk \citep[e.g.,][]{2020MNRAS.492.5030H}. That said, \citet{2017MNRAS.468L.108H} show that weak external radiation fields can launch winds without destroying CO and that such a mechanism could explain IM Lup's outer CO ``halo.'' Given that IM Lup and RU Lup lie within $10'$ of each other in the Lupus II cloud and are therefore presumably exposed to similar radiation fields, the two systems should be similarly susceptible to external photoevaporation. However, RU Lup's CO kinematics exhibit much stronger deviations from Keplerian rotation compared to IM Lup and thus would seemingly require a stronger radiation field to explain the perturbations. 

\subsubsection{Stellar or planetary mass companions}
Simulations indicate that stellar companions and massive ($>1$ $M_\text{Jup}$) planets can trigger spiral density waves in disks \citep[e.g.,][]{1986ApJ...307..395L, 2002ApJ...565.1257T}. RU Lup has no known stellar or planetary companions. Based on VLT NaCo imaging in $L'$ band, Jorquera et al. (submitted) estimate a $90\%$ detection probability for a  $\sim2$ $M_\text{Jup}$ planet beyond 200 au from RU Lup. No stars in the \textit{Gaia} catalog \citep{2018AA...616A...1G} are within $33''$ (5200 au in projection) of RU Lup. \textit{Gaia} has detected brown dwarfs in Lupus with masses as low as a few hundredths of a solar mass, but its mass sensitivity at any specific location in Lupus depends on the line-of-sight extinction as well as potential obscuration by circumstellar material \citep[e.g.,][]{2014AA...561A...2A, 2020AA...633A.114S}. Within $2'$ of RU Lup, the only \textit{Gaia} DR2 object with a relatively similar parallax is IRAS 15533-3742, an M3.5 T Tauri star coinciding in projection with the Herbig Haro object HH 55 \citep{1987ApJ...316..311C, 1990PASP..102..117H}. Prior to the detection of IRAS 15533-3742, RU Lup had been proposed as the energy source for HH 55 due to their proximity in projection  \citep{1977ApJS...35..161S}. $Gaia$ distance estimates now suggest that the two sources are likely well-separated, with RU Lup at $158.9\pm0.7$ pc and IRAS 15533-3742 at $145\pm5$ pc \citep{2018AJ....156...58B}. \citet{2007AA...461..253S} raised the possibility that RU Lup had a companion with a 3.7 day period based on radial velocity variations, but considered star spots to be the more likely explanation. Even if RU Lup has a close companion, we consider it unlikely to be the origin of RU Lup's spiral arms, which do not appear to extend further inward than $\sim260$ au from the star. 

Furthermore, RU Lup's spiral arm properties do not closely resemble models of spiral arms triggered by companions. Planetary-mass companions are only expected to induce second-order perturbations of the Keplerian velocity field rather than the dramatically non-Keplerian kinematics observed toward RU Lup \citep[e.g.,][]{2018MNRAS.480L..12P, 2018ApJ...860L..13P}. A stellar companion may still induce large-scale non-Keplerian spirals, as seen in $^{12}$CO emission around AS 205N \citep{2014ApJ...792...68S, 2018ApJ...869L..44K}. While a perturber in principle can trigger the formation of multiple spiral arms, simulations generally indicate that the spiral pattern will often have strong primary and/or secondary arms \citep[e.g.,][]{2015ApJ...809L...5D, 2018ApJ...859..119B, 2020MNRAS.491.1335R}, unlike what is seen in the RU Lup system.

\subsubsection{Stellar flybys}
Stellar flybys have been proposed to induce large-scale non-Keplerian spiral arms, increase stellar accretion rates, and lead to stellar brightness variations  \citep[e.g.,][]{1993MNRAS.261..190C, 2003ApJ...592..986P, 2020MNRAS.491..504C}. It is not clear from simulations whether clumping should be expected, but it is interesting to note that clumps are observed in $^{12}$CO emission around RW Aurigae, perhaps the most promising candidate for a system sculpted by a stellar flyby \citep{2018ApJ...859..150R}. Spirals induced by stellar encounters can have pitch angles as large as $30^\circ$ \citep{2019MNRAS.483.4114C}, comparable to the pitch angles of RU Lup's arms. However, stellar flyby simulations also generally only form one or two dominant spiral arms, sometimes with additional weaker arms \citep[e.g.,][]{2003ApJ...592..986P, 2005AJ....129.2481Q, 2020MNRAS.491..504C}. This behavior stands in contrast to RU Lup's multiple spiral arms with similar peak intensities (although intensity variations can be large within any given arm).

There is not yet an obvious candidate for a recent encounter with RU Lup. Since RU Lup is in a low-mass star-forming region, any encounter would most likely be with another star that has similar or lower mass. For equal-mass stellar encounters, \citet{2014AA...565A.130B} find that disks start to be significantly perturbed at periastron distances of a few hundred au. After periastron passage, spiral structures induced by the flyby are expected to survive for a few thousand years \citep{2019MNRAS.483.4114C}. The velocity dispersion of Lupus has been measured to be 1.3 km s$^{-1}$ \citep{2007ApJ...658..480M}. If we suppose that RU Lup underwent an encounter at a periastron distance of 500 au with a star at a fairly large relative velocity of 5 km s$^{-1}$, then the two stars would be separated by $\sim5800$ au after 5000 years. As noted in the previous discussion of potential stellar companions of RU Lup, the closest \textit{Gaia} DR2 source with a similar parallax is IRAS 15533-3742, which still has a projected separation of 19000 au. 
 
In summary, the complex CO emission features do not appear to be fully consistent with existing models of any of the aforementioned processes. RU Lup's large-scale spiral arms and complex CO kinematics may be evocative of models of stellar flybys, but the viability of this scenario is weakened by the lack of a stellar flyby candidate. RU Lup's CO behavior does not match closely with expectations for a bound companion (although the annular gaps in the millimeter continuum do suggest that planetary mass companions may orbit RU Lup). Evaluating whether a wind contributes to RU Lup's non-Keplerian kinematics will require identifying other processes responsible for the spiral arms and clumps and determining whether those processes are likely to co-exist with or launch a wind. While the large scale and irregularity of the spiral arms in conjunction with the presence of clumps are suggestive of gravitational instability, models of gravitational instability alone do not account for the observed kinematics. It is possible that more than one of the aforementioned processes could be occurring in RU Lup, thereby accounting for the full range of phenomena observed; for example, infall of cloud or envelope material can make a disk gravitationally unstable \citep[e.g.,][]{2005ApJ...633L.137V}.

\subsection{Comparison of RU Lup to other sources }
\subsubsection{Spiral arms in disks}
Spiral arms in molecular emission have only been reported for a small number of protoplanetary disks so far, including AB Aur in $^{12}$CO and possibly $^{13}$CO \citep{2005ApJ...622L.133C, 2006ApJ...645.1297L, 2012AA...547A..84T, 2017ApJ...840...32T}, HD 100453 in $^{12}$CO \citep{2020MNRAS.491.1335R}, HD 142527 in $^{12}$CO \citep{2014ApJ...785L..12C}, MWC 758 in $^{13}$CO \citep{2018ApJ...853..162B}, AS 205 in $^{12}$CO \citep{2018ApJ...869L..44K}, HL Tau in HCO$^+$ \citep{2019ApJ...880...69Y}, TW Hya in $^{12}$CO \citep{2019ApJ...884L..56T}, and UX Tau in $^{12}$CO \citep{2020ApJ...896..132Z}.  This small group exhibits diverse disk and spiral properties. Both Herbig Ae and T Tauri stars are represented. Similar to RU Lup, both the TW Hya and HL Tau disks have largely axisymmetric millimeter continuum emission featuring a series of narrow gaps and rings \citep{2015ApJ...808L...3A,2016ApJ...820L..40A}. Meanwhile, of the aforementioned sources, non-axisymmetric structures such as spiral arms and/or high-contrast crescent-like asymmetries have been detected in the millimeter continuum emission of HD 142527, AB Aur, MWC 758, and AS 205 \citep[e.g.,][]{2013Natur.493..191C, 2017ApJ...840...32T, 2018ApJ...860..124D, 2018ApJ...869L..41A, 2018ApJ...869L..44K}. Most of these systems only appear to have one or two arms, but AB Aur and AS 205 both appear to have at least four arms each and RU Lup has at least five arms. So far, systems with four or more arms all have large-scale non-Keplerian CO emission. Large-scale non-Keplerian molecular emission is only sometimes detected in systems with one or two spiral arms.

The properties of spiral arms detected in molecular emission are strikingly different from those detected in millimeter continuum. No more than two arms have been detected in millimeter continuum emission in any given disk \citep{2016Sci...353.1519P, 2018ApJ...860..124D, 2018ApJ...869L..43H, 2018ApJ...869L..44K, 2020NatAs...4..142L, 2020MNRAS.491.1335R}. Millimeter continuum spiral arms in T Tauri disks have so far all been roughly $180^\circ$ symmetric. In contrast, symmetric molecular spiral arms are atypical and have only been reported for the inner regions of the AB Aur disk \citep[e.g.,][]{2017ApJ...840...32T}. Because the millimeter continuum traces large dust grains in the midplane and molecular emission traces gas in the upper layers of disks, their different spiral morphologies may be in part a consequence of the disk vertical temperature gradient or of the decoupling of large dust grains from the gas \citep{2018MNRAS.474L..32J, 2019MNRAS.489.3758V}. Not all systems with millimeter continuum spiral arms have spiral arms observed in gas tracers and vice versa, perhaps pointing to multiple spiral arm formation mechanisms being active in disks.

\subsubsection{Disks surrounded by large-scale non-Keplerian CO emission}

RU Lup joins a growing number of detections of large-scale non-Keplerian, asymmetric structures in Class II systems, such as AB Aur \citep{2005ApJ...622L.133C, 2006ApJ...645.1297L, 2012AA...547A..84T}, AS 205 \citep{2014ApJ...792...68S, 2018ApJ...869L..44K}, DO Tau \citep{2020AJ....159..171F}, Elias 20 \citep{2018ApJ...869L..41A}, EX Lup \citep{2018ApJ...859..111H}, FS Tau A \citep{2020ApJ...889..140Y}, J15450634-3417378 \citep{2018ApJ...859...21A}, RW Aur \citep{2018ApJ...859..150R}, SU Aur \citep{2019AJ....157..165A}, UX Tau \citep{2020ApJ...896..132Z} and WSB 52 \citep{2018ApJ...869L..41A}. A few members of this group overlap with the spiral-armed systems discussed above, but other morphologies include collimated tails, arcs, clumps, shells, and broad asymmetries. It is not clear the extent to which these complex morphologies have a common origin. In at least a few cases (AS 205, FS Tau A, RW Aur, UX Tau), the complex gas behavior appears to be the consequence of dynamical interaction between multiple stars. In other cases, large-scale gas structures around Class II disks may be related to interaction with a remnant protostellar envelope. Millimeter continuum and molecular line observations of FU Ori sources such as V883 Ori \citep{2016Natur.535..258C, 2017MNRAS.468.3266R} and borderline Class I/Class II sources such as HL Tau and DG Tau \citep{2015ApJ...808L...3A, 2020AA...634L..12D} have demonstrated that it is possible to form and maintain well-organized dust disks surrounded by infalling and/or outflowing gas. 

\subsubsection{Similarities with FU Ori systems}

RU Lup's CO morphology, elevated accretion rate, and photometric variability (described in more detail earlier in Section \ref{sec:GI}) suggest some kinship with outbursting FU Ori sources. Observationally, FU Ori sources are protostellar systems typically characterized by a rapid brightening event of up to several magnitudes at optical wavelengths, high accretion rates up to $10^{-4}$ $M_\odot$ yr$^{-1}$, and ejection of material through winds \citep[e.g.,][]{1996ARAA..34..207H}. Polarimetric images and interferometric CO observations of FU Ori sources show outflows, tails, and arms extending out to thousands of au \citep[e.g.,][]{2016SciA....2E0875L, 2017MNRAS.465..834Z, 2017MNRAS.466.3519R, 2018ApJ...864...20T}. FU Ori outbursts are generally associated with Class 0/I disks still deeply embedded in envelopes \citep[e.g.,][]{1996ARAA..34..207H}. The outbursts are most commonly attributed to infall from either the envelope or from a gravitationally unstable disk \citep[e.g.,][]{1996ARAA..34..207H, 2005ApJ...633L.137V}. \citet{2018ApJ...859..111H} have previously suggested that scaled-down versions of FU Ori events occur in EX Lup, a Class II disk that has undergone several moderate brightening events and exhibits an FU Ori-like outflow structure traced by CO emission. Thus, sources such as EX Lup and RU Lup may represent an evolutionary link between young, extremely active FUors and the more evolved, comparatively quiescent classical T Tauri stars. Other proposed FU Ori mechanisms suggest that outbursts occur due to environmental triggers rather than as part of an evolutionary stage in star formation. Such environmental triggers could include stellar flybys \citep[e.g.,][]{2008AA...487L..45P} and accretion of material from cloud fragments \citep[e.g.,][]{2019AA...628A..20D}, which were both discussed in the previous section as possible explanations for RU Lup's complex CO behavior. 

\subsubsection{Spiral arms in other systems}
RU Lup's CO emission is also reminiscent of spiral arms in sources that are not protostellar or protoplanetary disks. It bears a striking resemblance to the CO emission of EP Aquarii, an AGB star with  long and clumpy spiral arms, a clumpy ring interior to the spiral arms, and crescent-like emission interior to the ring \citep{2018AA...616A..34H}. \citet{2018AA...616A..34H} hypothesize that EP Aquarii's CO emission is sculpted by a companion interacting with the AGB wind. Clumpy spiral arms have also been observed in a large number of spiral galaxies in optical and UV images \citep[e.g.,][]{2004ApJ...604L..21E, 2015ApJ...800...39G}. These clumpy spiral galaxies have been hypothesized to form via disk instabilities \citep[e.g.,][]{2009ApJ...703..785D} or mergers \citep{2010MNRAS.406..535P}. While the detailed physics of AGB stars and galaxies obviously differ from those of protoplanetary disks, the morphological similarities motivate the exploration of analogous processes during star formation.  

\section{Summary\label{sec:summary}}
We present deep ALMA observations of the $J=2-1$ transition of $^{12}$CO, $^{13}$CO, and C$^{18}$O as well as the $J=3-2$ transition of DCO$^+$ toward the T Tauri star RU Lup in order to probe the gas distribution and kinematics. Our key findings are as follows: 

\begin{enumerate}
\item The CO isotopologues together trace four components of the RU Lup system: a compact Keplerian disk with a radius of $\sim120$ au, a non-Keplerian ``envelope-like'' structure surrounding the Keplerian disk, at least five spiral arms extending up to 1000 au from RU Lup, and at least eleven clumps outside the spiral arms. 
\item The spiral arms are clumpy, exhibiting localized intensity variations of at least a factor of 3. Their inward termination point appears to be at a clumpy CO emission ring that is $\sim260$ au from RU Lup. The spirals are quite open, with pitch angles up to 30$^\circ$. They are generally blueshifted relative to the systemic velocity.
\item The clumps outside the spiral arms are found at a range of velocities, but most are redshifted relative to the systemic velocity. Most of the clumps do not appear to be bound to RU Lup. The peak $^{12}$CO brightness temperatures indicate the clumps could be optically thick, rendering mass estimates uncertain. Based on the $^{12}$CO emission and non-detection in continuum emission, we estimate that the clump masses are likely between $\sim0.1-150$ M$_\earth$. Other major sources of uncertainty that could change this range include the $^{12}$CO:H$_2$ ratio and the gas-to-dust ratio. 
\item RU Lup's complex CO emission, enhanced stellar accretion, and strong photometric variability are reminiscent of a scaled-down version of FU Ori outburst behavior. Mechanisms that have been proposed for the FU Ori phenomena include gravitational instability, delivery of material from envelopes or cloud fragments, or stellar flybys. One of these explanations may also be applicable to RU Lup. 
\end{enumerate}

An increasing number of Class II systems are now known to be surrounded by large-scale, complex gas structures rather than simply axisymmetric Keplerian disks. This raises the question of how common these complex structures are. The available constraints are surprisingly mediocre. Surveys of star-forming regions have usually been shallow, leading to a large fraction of non-detections of $^{12}$CO even with ALMA \citep[e.g.,][]{2016ApJ...827..142B, 2018ApJ...859...21A}. A few surveys have only covered the $^{13}$CO and C$^{18}$O isotopologues, which are generally only abundant enough to trace the inner regions of protoplanetary disks \citep[e.g.,][]{2017ApJ...844...99L, 2018ApJ...869...17L}. A deep, comprehensive survey of gas structures around young stars will be key for establishing the typical conditions of planet formation. 

\acknowledgments
We thank the anonymous referee for helpful comments improving this manuscript. We also thank Henning Avenhaus for sharing his SPHERE images of RU Lup and Stefano Facchini, Dimitar Sasselov, Mercedes Lopez-Morales, and Fred Ciesla for helpful discussions. This paper makes use of ALMA data\\
\dataset[ADS/JAO.ALMA\#2018.1.01201.S]{https://almascience.nrao.edu/aq/?project\_code=2018.1.01201.S} and \\
\dataset[ADS/JAO.ALMA\#2016.1.00484.L]{https://almascience.nrao.edu/aq/?project\_code=2016.1.00484.L}. We thank NAASC and JAO staff for their advice on data calibration and reduction. ALMA is a partnership of ESO (representing its member states), NSF (USA) and NINS (Japan), together with NRC (Canada) and NSC and ASIAA (Taiwan), in cooperation with the Republic of Chile. The Joint ALMA Observatory is operated by ESO, AUI/NRAO and NAOJ. The National Radio Astronomy Observatory is a facility of the National Science Foundation operated under cooperative agreement by Associated Universities, Inc. J. H. acknowledges support from the National Science Foundation Graduate Research Fellowship under Grant No. DGE-1144152.  S. A. and J. H. acknowledge funding support from the National Aeronautics and Space Administration under Grant No. 17-XRP17\_2-0012 issued through the Exoplanets Research Program. L. P. acknowledges support from CONICYT project Basal AFB-170002 and from FONDECYT Iniciaci\'on project \#11181068. This research has made use of NASA's Astrophysics Data System.

\facilities{ALMA, VLT:Melipal}

\software{ \texttt{analysisUtils} (\url{https://casaguides.nrao.edu/index.php/Analysis_Utilities}), \texttt{AstroPy} \citep{2013AA...558A..33A}, \texttt{CASA} \citep{2007ASPC..376..127M}, \texttt{dsharp\_opac} \citep{2018ApJ...869L..45B}, \texttt{emcee} \citep{2013PASP..125..306F}, \texttt{matplotlib} \citep{Hunter:2007}, \texttt{SciPy} \citep{scipy}}

\appendix

\section{Properties of continuum sources near RU Lup \label{sec:contsources}}
\begin{figure}
\begin{center}
\includegraphics{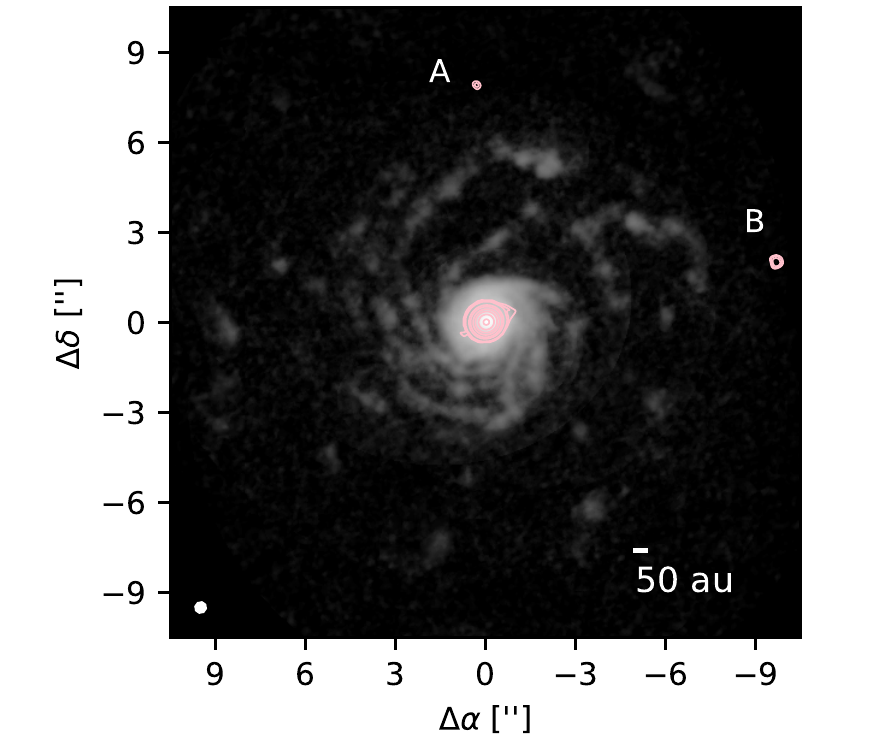}
\end{center}
\caption{1.3 mm continuum emission (pink contours) detected in the field of RU Lup, drawn on top of the $^{12}$CO emission (greyscale). Contours are drawn at [5, 7, 10, 50, 100, 250, 500, 1000, 2000]$\sigma$, where $\sigma=0.03$ mJy beam$^{-1}$. The continuum rms is measured inside an annulus with an inner radius of $9''$ and outer radius of $10''$.  The central object is the RU Lup disk. The asymmetry in its $5\sigma$ contour is due to PSF artifacts, since the RU Lup disk is extremely bright (the peak intensity at 1.3 mm is 70.85 mJy).  \label{fig:otherdustsources}}
\end{figure}

Two faint, compact continuum sources are visible in the same field as RU Lup (Figure \ref{fig:otherdustsources}). We assign the label ``Source A'' to the one north of RU Lup and the label ``Source B'' to the one west of RU Lup. To our knowledge, these sources have not previously been described in the literature. To measure the positions and fluxes of these sources, we fit two-dimensional Gaussians using the \texttt{imfit} task in CASA and list the results in Table \ref{tab:contproperties}. 

Both sources lie outside RU Lup's CO spiral arms, and neither coincides with the positions of any of the CO clumps. No object in the \textit{Gaia} DR2 catalog \citep{2018AA...616A...1G} corresponds to the positions of either Source A or B. The two sources can plausibly be attributed to background sub-millimeter galaxies. Using the Schecter function derived in \citet{2015AA...584A..78C} from ALMA continuum maps and assuming that galaxies are uniformly distributed, the probability that a sub-millimeter galaxy with a 1.3 mm flux density exceeding $0.31$ mJy is found within $10''$ of RU Lup is $\sim16\%$. Likewise, the probability of a source with a flux density exceeding $0.71$ mJy in this region is $\sim4\%$. Measuring the SED and proper motion of these sources will be useful for confirming their nature.

\begin{deluxetable}{ccc}
\tablecaption{Continuum Source Properties\label{tab:contproperties}}
\tablehead{
\colhead{Source}&\colhead{Position relative to RU Lup}&\colhead{Flux\tablenotemark{a}}\\
\colhead{} & \colhead{(arcsec, arcsec)} & \colhead{(mJy)}}
\startdata
A&($0.32\pm0.02$, $7.9\pm0.02$)&$0.31\pm0.06$\\
B& ($-9.67\pm0.01$, $2.00\pm0.01$)&$0.71\pm0.08$
\enddata
\tablenotetext{a}{Uncertainties quoted in the table are statistical. The systematic flux calibration uncertainty contributes another $10\%$.}
\end{deluxetable}

\section{CLEAN mask for $^{12}$CO\label{sec:maskdescription}}

For channels between $-5.5$ and $0.75$ km s$^{-1}$, a circular mask with a radius of $1\farcs2$ was used to CLEAN the emission. The starting CLEAN mask for the remaining channels is illustrated in Figure \ref{fig:COmask}. 

\begin{figure*}
\begin{center}
\includegraphics{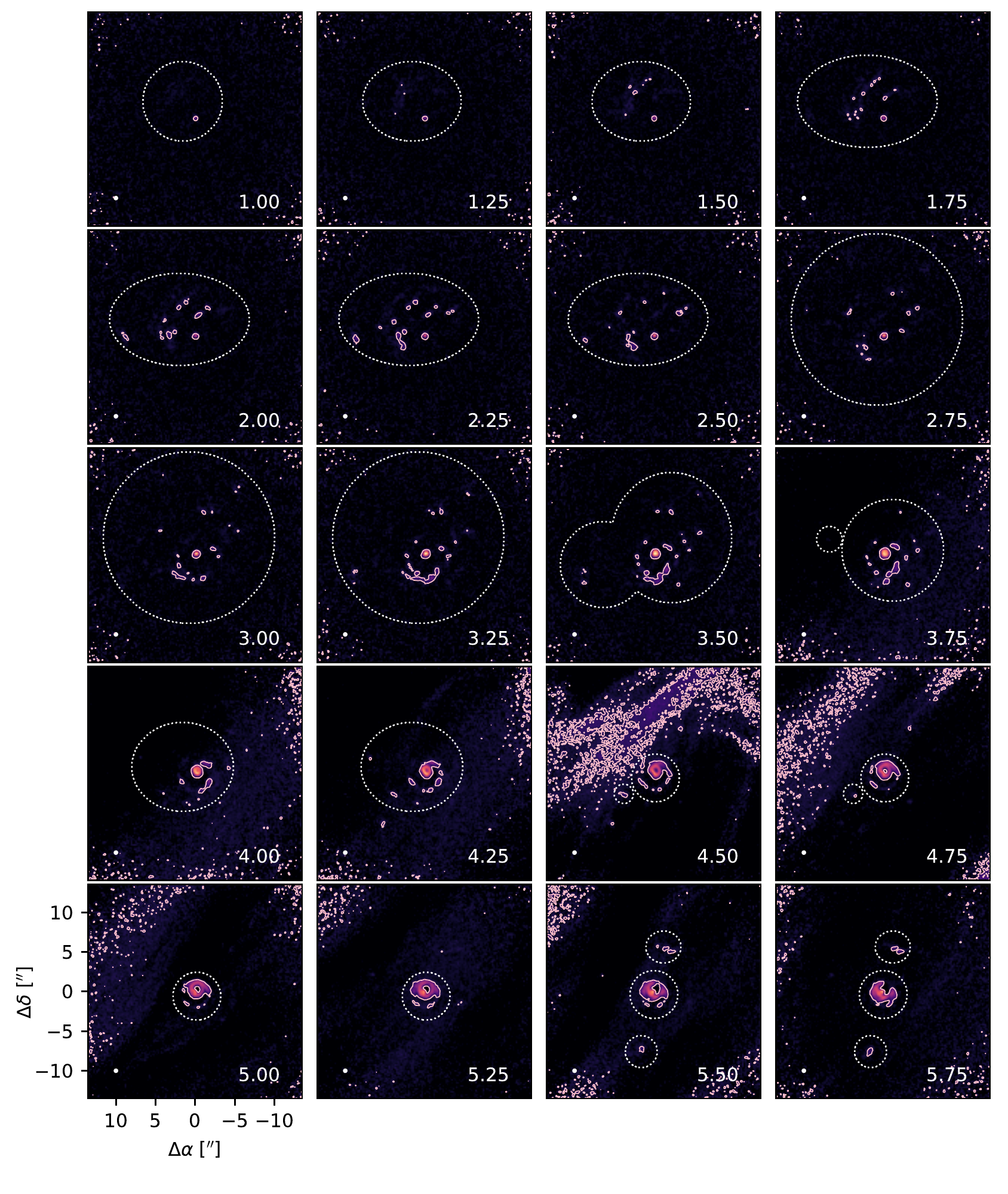}
\end{center}
\caption{CLEAN mask used for imaging $^{12}$CO overlaid on the line channel maps. The dotted white ellipses show the mask boundaries. The pink contours mark emission at the 12.5 mJy beam$^{-1}$ level (equal to $5\sigma$ at a distance of $\sim10''$ from the phase center). Cloud contamination is visible in channels from 3.75 to 6.5 km s$^{-1}$. The corners of each image are noisy due to the decreased sensitivity near the edge of the primary beam. The synthesized beam is drawn in the lower left corner of each panel. The LSRK velocity (km s$^{-1}$) is labeled in the lower right corner. Offsets from the phase center (in arcseconds) are marked on the axes in the lower left corner of the figure. \label{fig:COmask}}
\end{figure*}

\begin{figure*}
\begin{center}
\ContinuedFloat
\includegraphics{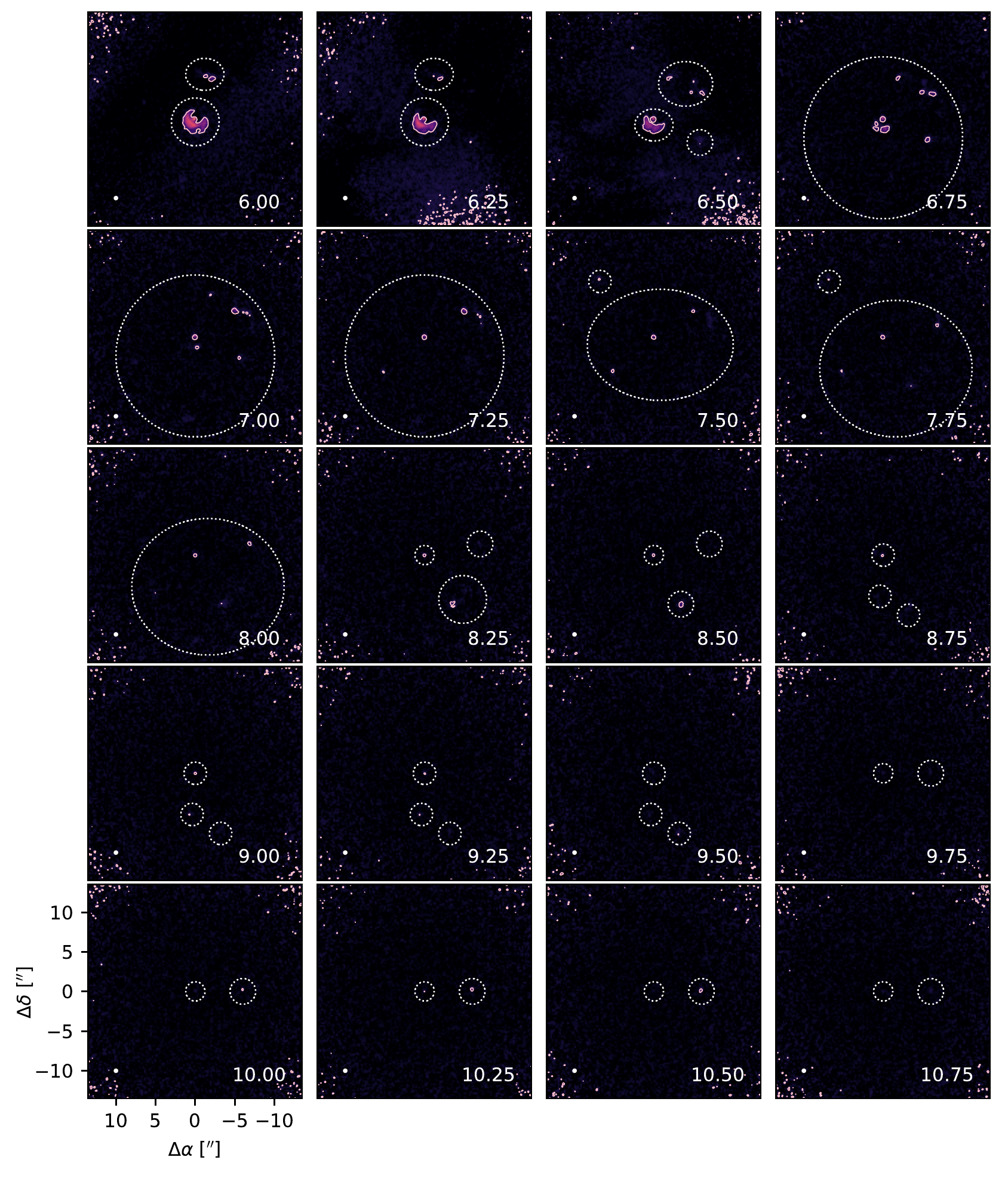}
\end{center}
\caption{Continued}
\end{figure*}

\end{document}